\documentclass[a4paper,nofootinbib,showpacs,aps,floatfix,superscriptaddress]{revtex4}
\usepackage{graphicx}        
\usepackage{amsmath}
\usepackage{bm}

\newcommand{\ii}{\mathrm{i}}                  
\newcommand{\e}{\mathrm{e}}                  
\newcommand{\ket}[1]{|#1\rangle}             
\newcommand{\bra}[1]{\langle #1|}            
\newcommand{\braket}[2]{\langle #1|#2\rangle}
\renewcommand{\d}[1]{\mathrm{d}#1} 
\newcommand{\tr}{\mathrm{Tr}}                
\renewcommand{\Re}{\mathrm{Re}}              
\newcommand{\mt}[1]{\mathrm{#1}}             
\newcommand{\la}{\langle}
\newcommand{\ra}{\rangle}
\newcommand{\wh}[1]{\widehat{#1}}
\newcommand{\be}{\begin{equation}}
\newcommand{\ee}{\end{equation}}
\newcommand{\beqa}{\begin{eqnarray}}
\newcommand{\eeqa}{\end{eqnarray}}
\newcommand{\beq}{\begin{equation}}
\newcommand{\eeq}{\end{equation}}

\begin{document} 
\title{Dwell-time distributions in quantum mechanics}
\author{J. Mu\~noz}
\email{josemunoz@saitec.es}
\affiliation{Dpto. de Qu\'\i mica F\'\i sica, UPV-EHU, 
Apdo. 644, 48080 Bilbao, Spain}
\author{I. L. Egusquiza}
\email{inigo.egusquiza@ehu.es}
\affiliation{Dept. of Theoretical Physics, The University of the Basque Country,
Apdo. 644, 48080 Bilbao, Spain}
\author{A. del Campo}
\email{a.del-campo@imperial.ac.uk}
\affiliation{Institute for Mathematical Sciences, Imperial College London, SW7 2PE, UK; QOLS, The Blackett Laboratory, Imperial College London, Prince Consort Rd., SW7 2BW, UK}
\author{D. Seidel}
\email{dirk_x_seidel@yahoo.de}
\affiliation{Dpto. de Qu\'\i mica F\'\i sica, UPV-EHU, 
Apdo. 644, 48080 Bilbao, Spain}
\author{J. G. Muga}
\email{jg.muga@ehu.es}
\affiliation{Dpto. de Qu\'\i mica F\'\i sica, UPV-EHU, 
Apdo. 644, 48080 Bilbao, Spain}

\begin{abstract}
Some fundamental and formal aspects of the quantum dwell time are reviewed,
examples for free motion and scattering off a potential barrier are provided, as well as extensions of the concept. 
We also  examine the connection between the dwell time of a quantum particle 
in a region of space
and flux-flux correlations at the boundaries, as well as  
operational approaches and approximations to measure the
flux-flux correlation function and thus the second moment of the dwell time,
which is shown to be characteristically quantum, and larger than the corresponding classical moment even for freely moving 
particles.
\end{abstract}
\maketitle

\section{Introduction}

Time observables \index{time observables} 
in quantum mechanics have a long and debated history
\cite{Muga-book}. In spite of the fact that 
random time variables, measured after a system is prepared,
are common in laboratories, most often it has been argued that questions
about time in quantum mechanics should best be left alone, as illustrated
by the frequent reference to Pauli's theorem\index{Pauli's theorem}. Alternatively,
the emphasis has been laid on characteristic times, i.e. single time
quantities characterizing a process such as tunneling, or decay. This,
in many ways, runs counter to the usual procedure in quantum mechanics,
where additionally to the average value of a quantity we 
require prediction of higher order moments of that quantity; in other
words, the probability distribution.
 
 
With regard
to the time-of-arrival \index{time of arrival} observable several such distributions have
been proposed and studied, see volume 1 of ``Time in Quantum Mechanics'' \cite{toabook}, or \cite{MuLea-PR-2000}.  
In this chapter 
we analyze the dwell time,\index{dwell time} which appears to be a much simpler time
observable because the associated operator will indeed
be self-adjoint (over the adequate domain). 
At first sight, it could be thought
that this statement contradicts Pauli's theorem\index{Pauli's theorem}, which asserts
that no self-adjoint time observable can exist with canonical commutation
relations with a semi bound Hamiltonian. However, the dwell time 
is an \textit{interval} quantity, as opposed to the \emph{instant}
quantity that the time of arrival describes. Its associated operator,
therefore, should \emph{commute} with the Hamiltonian, as opposed
to presenting canonical commutation with it.

The dwell time of a particle in a region of space and its close relative, 
the delay time\index{delay time} \cite{Smith60}, are rather fundamental quantities 
that characterize the duration
of collision processes, the lifetime of unstable systems \cite{EkSie-AP-1971},
the response to perturbations \cite{JaWa-PRA-1989},
ac-conductance in mesoscopic conductors \cite{Bu},
or the properties of chaotic scattering \cite{MW}. 
In addition, the importance of dwell and delay times is underlined by 
their relation to the density of states, 
and to the virial expansion in statistical mechanics \cite{Nuss2002}.
We could thus hardly fail to study and characterize in detail such a
prominent quantity. 
For a sample of theoretical studies on the quantum dwell time see
\cite{EkSie-AP-1971,JaWa-PRA-1989, SK91,BSM94,LM94,Nuss2002,S02,DEMN04,Y04,S04,LV04,W06,K07,S07,BS08,charac,chokoloki}. A recurrent topic has been its role and decomposition in tunneling collisions. Instead, we shall focus here on a different, so far overlooked but fundamental aspect, namely,
the measurability and physical implications of its distribution, and its second moment.

Despite the nice properties of the dwell-time operator,   
the relevance of the concept and average value in many different fields,
or the apparent formal simplicity stated above, 
the dwell time is actually rather subtle and remains elusive and challenging in many ways. 
In particular, a     
direct and sufficiently non-invasive measurement, so that the statistical moments are produced by averaging over  
individual dwell-time values, is yet to be discovered. If the particle is detected (and thus localized) 
at the entrance of the region of interest,
its wavefunction is severely modified (``collapsed''),
so that the times elapsed 
until a further detection when it leaves the region do not reproduce 
the ideal dwell-time operator distribution, and depend on the 
details of the localization method. Proposals for operational, i.e.,
measurement-based approaches to traversal times based on model detectors which study the effect of localization 
have been discussed by Palao et al. \cite{PaMuBrJa-PLA-1997} and by Ruschhaupt \cite{Ru-PLA-1998}. For attempts to measure the dwell time with continuous or kicked ``clocks'' coupled to the particle's presence in the region see \cite{chokoloki,ASM93}.
All operational approaches to the quantum dwell time known so far 
have provided
only its average, and indirectly, by deducing it     
from its theoretical relation to some other observable with measurable average.
The average is obtained for example by a ``Larmor clock", using a weak homogeneous magnetic field in the region $D$ and the amount of spin rotations of an incident spin-$\frac{1}{2}$ particle \cite{Baz-SJNP-1967, Ry-SJNP-1967, Bue-PRB-1983}. An optical analogue is provided by the ``Rabi clock" \cite{Bra-JPB-1997}. 
It can also be deduced from average passage times  
at the region boundaries \cite{charac}, 
as well as by measuring the total absorption if a weak complex absorbing potential acts in the region \cite{CP1,CP2,CP3}. This last setup could be implemented with 
cold atoms and lasers as described in \cite{DEMN04,CP4} and will be discussed in 
Sect. \ref{sec6}.   
            

The chapter is organized as follows: the first sections are devoted to
fundamental and formal aspects (Sect. 2), examples (Sects. 3 and 4),
or extensions (Sect. 5) 
of the dwell-time concept and operator, whereas Sect. 6 tackles the relationship between the moments of the dwell-time operator and flux-flux correlation 
functions (ffcf) \cite{Munoz}, generalizing an approach by Pollak and Miller \cite{PoMi-PRL-1984}.  
They showed that the average stationary dwell time agrees with the first moment of a microcanonical ffcf. We shall see that this relation holds also for the second moment, but not for higher moments, and extend their analysis to the time-dependent (wavepacket) case. We shall also discuss a possible scheme to measure ffcf's, thus paving the way towards experimental access to quantum 
features of the dwell-time distribution.
\section{The Dwell Time Operator}
Unlike other time quantities, there has been 
a broad consensus on the operator representation 
of the dwell time \cite{EkSie-AP-1971, JaWa-PRA-1989}.
For one particle evolving
unitarily with Hamiltonian $\wh{H}$ in region $D$,
which we limit here to one dimension for simplicity,
$D= \{x: x_1 \leq x \leq x_2\}$, it takes the form 
\beqa
\widehat{T}_{D}&=&\int_{-\infty}^\infty \d t\,\wh{\chi}_D(t)
=\int_{-\infty}^{\infty}\d t\, \e^{\ii\wh{H}t/\hbar}\chi_{D}(\wh{x})\, 
\e^{-\ii\wh{H}t/\hbar}\;,
\label{eq:definition}
\eeqa
where 
$\wh{\chi}_D(t)$ is the (Heisenberg) projector onto $D$,     
%
%
and $\chi_D(\wh{x})=\wh{\chi}_D(0)=\int_{x_1}^{x_2} dx|x\ra\la x|$. 
Without delving too far in the functional analysis definition, i.e.
into the proper description of its domain and of its adjoint, we can
see that $\wh{T}_D$ will be self-adjoint (we shall come back to
this issue after studying the specific example of the free particle
Hamiltonian in Sect. \ref{sec:free}).\index{dwell time operator} 
At any rate, it
is clear that, at least formally, this operator commutes with the
Hamiltonian, as can be seen from what follows:\footnote{Time-limited 
versions of the dwell-time operator such as $\int_0^\infty \d t\,\e^{\ii\wh{H}t/\hbar}\chi_{D}(\wh{x})\, \e^{-\ii\wh{H}t/\hbar}$ do {\it not} generally commute with
$\wh{H}$, see \cite{EkSie-AP-1971} or Sect. \ref{tor}.}    
\begin{eqnarray}
\widehat{T}_{D}\e^{-\ii\wh{H}t/\hbar} & = & \int_{-\infty}^{\infty}\d \tau\, \e^{\ii\wh{H}\tau/\hbar}\chi_{D}(\wh{x})\, \e^{-\ii\wh{H}(\tau+t)/\hbar}\\
 & = & \int_{-\infty}^{\infty}\d \tau\, \e^{\ii\wh{H(}\tau-t)/\hbar}\chi_{D}(\wh{x})\, \e^{-\ii\wh{H}\tau/\hbar} =  \e^{-\ii\wh{H}t/\hbar}\widehat{T}_{D}\,\,.\nonumber
\end{eqnarray}
The commutation of $\widehat{T}_{D}$ and the Hamiltonian leads us to
search for the eigenfunctions of dwell time in the stationary eigenspaces
of the latter. Let $\alpha$ be the degeneracy index for these eigenspaces,
such that $\wh{H}|E,\alpha\rangle=E|E,\alpha\rangle$. We easily
obtain the matrix elements of $\widehat{T}_{D}$ in the corresponding
eigenspace, 
\begin{equation}
\widehat{T}_{D}|E,\alpha\rangle=2\pi\hbar\sum_{\beta}\langle E,\beta|\chi_{D}(\wh{x})|E,\alpha\rangle\,|E,\beta\rangle\;.
\end{equation}
We may thus reduce the problem of eigenvalues and eigenfunctions
of the dwell-time operator to a set of matrix diagonalization problems
in each of the eigenspaces of the Hamiltonian.

Except in Sect. \ref{sec5} we shall assume that the Hamiltonian holds a purely continuous spectrum
with degenerate (delta-normalized) scattering eigenfunctions $|\phi_{\pm k}\ra$ corresponding to 
incident plane waves $|\pm k\ra$, with energy $E=k^2\hbar^2/(2m)$, normalized as 
$\la k|k'\ra=\la \phi_k|\phi_{k'}\ra=\delta(k-k')$.  
  
Following the same manipulation done for the $S$-operator in one-dimensional 
scattering theory \cite{charac},  
it is convenient to define an on-the-energy-shell
$2\times 2$ dwell-time matrix $\mathsf{T}$, by factoring out an energy delta,  
\beq
\la \phi_k|\wh{T}_D|\phi_k'\ra=\delta(E-E')\frac{|k|\hbar^2}{m}
\mathsf{T}_{kk'}\;, 
\label{dele} 
\eeq
where $E'=k'^2\hbar^2/(2m)$ and   
\beq
\mathsf{T}_{kk'}=
\la \phi_k|{\chi}_D(\wh{x})|\phi_{k'}\ra\frac{2\pi m}{|k|\hbar},\;\; 
E=E'\;.
\eeq
In particular, $\mathsf{T}_{kk}$ is 
the average dwell time for a finite space region defined by B\"uttiker in the stationary regime \cite{Bue-PRB-1983},
\be \label{eq:dwell_stat}
\mathsf{T}_{kk}= \frac{1}{|j(k)|} \int_{x_1}^{x_2} \d x\, |\phi_k(x)|^2\;,
\ee
where $j(k)$ is the incoming flux associated with $\ket{\phi_k}$. 

An intriguing peculiarity of the quantum dwell time is that the diagonalization of $\mathsf{T}$ at a given energy provides in general two distinct 
eigenvalues $t_\pm(k)$, $k>0$, 
and corresponding eigenvectors $|t_\pm(k)\ra$, even in cases in which 
only a single 
classical time exists, such as free motion, or transmission
above the barrier; some explicit examples are discussed below.  
A consequence is a broader variance of the quantum dwell-time distribution 
compared to the classical one. 

The quantum dwell-time distribution\index{dwell time distribution} for a state $|\psi\ra=|\psi(t=0)\ra$, is formally given by 
\be \label{eq:dwell_distri}
\Pi_\psi(\tau) = \bra{\psi} \delta(\wh{T}_D -\tau) \ket{\psi}\;, 
\ee
since the self-adjointness of the dwell-time operator implies that the spectral
theorem applies, and that operator moments coincide with the moments
of the distribution. 
We shall not consider in this work other sources of fluctuations such as mixed states or ensembles of Hamiltonians. In these two cases one could compute distributions of {\it average} dwell times, whereas here we shall be interested in the 
distribution of the dwell time itself, but only for pure states
and a single Hamiltonian.

On computing the distribution of dwell times we run into the difficulty,
mentioned above, that the dwell-time operator is multiply degenerate,
and that it is not altogether easy to de-scramble the relation between
a given $t$ and the corresponding set of momenta. It is useful
in this case to profit from the spectral theorem to compute the generating
function of the dwell-time distribution, defined as 
\begin{eqnarray}
f_{\psi}(\omega)=\int_{0}^{\infty}\d t\, \e^{\ii\omega t}\,\Pi_{\psi}(t)
=\langle\psi|\e^{\ii\omega\widehat{T}_{D}}|\psi\rangle\;.
\end{eqnarray}
Normalizing the dwell-time eigenvectors 
so as to have the resolution of the identity  
\beq
\label{one}
\widehat{1}=\sum_\alpha \int_0^\infty \d k |t_\alpha(k)\ra\la t_\alpha(k)|\;,
\eeq
$f_\psi$ can be written as
\beqa
f_{\psi}(\omega)  =  \int_{0}^{\infty}\d k\,\Big[\e^{\ii\omega t_{+}(k)}\langle\psi|t_{+}(k)\rangle\langle t_{+}(k)|\psi\rangle
+\e^{\ii\omega t_{-}(k)}\langle\psi|t_{-}(k)\rangle\langle t_{-}(k)|\psi\rangle\Big]\;,
\label{fomega}
\eeqa
and, hence,
\beqa
\Pi_{\psi}(t)&=&\int_{-\infty}^{\infty}\frac{d\omega}{2\pi}\, \e^{-\ii\omega t}f_{\psi}(\omega)
\nonumber\\
&=&\int_0^\infty \d k
\left[\delta(t-t_+) |\la\psi|t_+\ra|^2+\delta(t-t_-)|\la \psi|t_-\ra|^2\right]\;.
\label{eq:distfree}
\eeqa

The average of the dwell time for a wavepacket  
can be given in terms of the position probability density  
in correspondence (unlike its second moment, as shown below) to the 
expression for an ensemble of classical particles. It reads \cite{JaWa-PRA-1988, MuBrSa-PLA-1992} 
\beqa
\label{taud}
\la \psi|\wh{T}_D|\psi\ra=
\int_{-\infty}^\infty \d t \int_{x_1}^{x_2} \d x\,|\psi(x,t)|^2
=\int_0^\infty \d k\,|\la k|{\psi}^{in}\ra|^2 \mathsf{T}_{kk}\;,
\eeqa
where $\psi(x,t) = \int_0^\infty \d k\, \la\phi_k|\psi\ra \exp(-\ii\hbar k^2 t/2m) \phi_k(x)$ is the time-dependent wave packet and we assume, here and in the rest of the Chapter, incident 
wavepackets with positive momentum components. 
To write Eq. (\ref{taud}) use has been made of the standard scattering relation 
$\la \phi_k|\psi\ra=\la k|\psi^{in}\ra$, where  
$\la x|k\ra=(2\pi)^{-{1/2}}\exp(\ii kx)$, and $\psi^{in}$ is the freely-moving
asymptotic incoming state of $\psi$. 
Space-time integrals of the form (\ref{taud}) had been used
to define time delays by comparing the free motion to that with a scattering center and taking the limit of infinite volume \cite{GoWa-book}. 

The second moment takes the form, as before for 
wavepackets with incident positive momentum components, 
\beqa
\left\langle \widehat{T}_D^2\right\rangle
&=&
\int_0^\infty \d k (\mathsf{T}^{\,2})_{kk}|\la k|\psi^{in}\ra|^2=\int_0^\infty \d k (|\mathsf{T}_{kk}|^2+|\mathsf{T}_{k-k}|^2)|\la k|\psi^{in}\ra|^2
\nonumber\\
&=&\int_0^\infty \d k \frac{4\pi^2 m^2}{\hbar^2 k^2}
\Bigg[\Big(\int_{x_1}^{x_2} \d x\, |\phi_k(x)|^2 \Big)^2
+\Big| \int_{x_1}^{x_2}\! \d x\, \phi_k^*(x) \phi_{-k}(x) \Big|^2 \Bigg]|\la k|\psi^{in}\ra|^2
\;, 
\eeqa
with a term without classical counterpart. 
\section{The Free Particle Case\label{sec:free}}
\begin{figure}
\includegraphics[scale=.23]{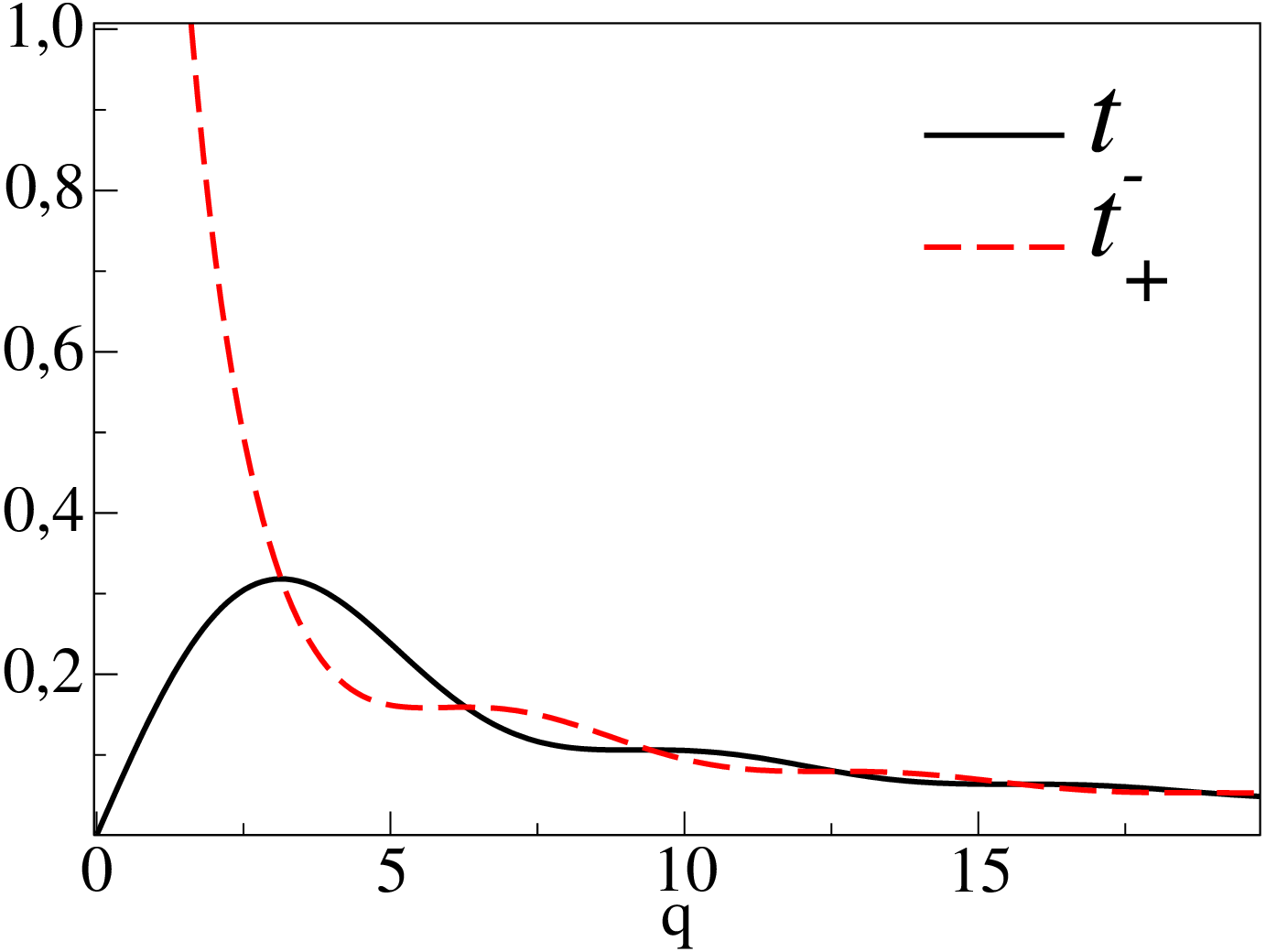}
\caption[]{Eigenvalues of the dwell-time-operator as a function of $q=kL$, in units $mL^2/\hbar$, for a freely moving particle in a region of length $L$}
\label{figfree}
\end{figure}

In order to get a better grasp of the abstract results,
it is adequate to illustrate them with the
explicitly solvable free-particle case for a region $D$ that extends
from $x_1=0$ to $x_2=L$. Indeed the humble freely moving particle turns
out to be rather interesting and surprisingly complex with regard to the 
dwell time. 
The free-particle Hamiltonian is doubly
degenerate; we can choose the degeneracy indices to coincide with
the sign of the momentum, and, on computation, with due attention
to the different normalization of the energy and the momentum eigenfunctions,
we find the following generalized eigenfunctions of the free dwell-time
operator\index{dwell-time, free motion}:
\begin{equation}
|t_{\pm}(k),D\rangle=\frac{1}{\sqrt{2}}\left[|k\rangle\pm \e^{\ii kL}|-k\rangle\right]\;,
\label{tvec}
\end{equation}
where
\begin{equation}
t_{\pm}(k)=\frac{mL}{k\hbar}\left(1\pm\frac{1}{kL}\sin{kL}\right)\label{eq:freeeigenvalue}
\end{equation}
are the corresponding eigenvalues, 
in clear contrast to the classical time $t_{\mathrm{class}}=mL/(|k|\hbar)$, 
see Fig \ref{figfree}.
It is important to notice that
the inverse function is multivalued.
Due to this multivaluedness we have kept generalized
eigenfunctions with the dimensionality of $|k\rangle$, instead of
normalizing them to the delta function in dwell times.
With this normalization it is easy to check that the resolution of the identity 
has the form (\ref{one}). 
%
%

Interestingly, $t_-(k)$ tends to zero as $k\to 0$, one more very non-classical 
effect that is better understood from the coordinate representation of the 
eigenvectors, 
\beqa
\la x|t_+\ra&=&\frac{\e^{\ii kL/2}}{\pi^{1/2}}\cos[k(x-L/2)]\;,
\nonumber\\
\la x|t_-\ra&=&\frac{\ii\e^{\ii kL/2}}{\pi^{1/2}}\sin[k(x-L/2)]\;,
\eeqa
symmetric and antisymmetric with respect to the interval center.  
$\la x|t_-\ra$ vanishes at $L/2$, 
so the particle density tends to vanish in $D$ 
with longer wavelengths as $k\to 0$.  
 
The dwell-time operator for an interval $D'=[a,a+L]$ is obtained
from the one above by translation,
\begin{equation}
\widehat{T}_{D\,'}=\e^{-\ii a\wh{k}}\widehat{T}_{D}\e^{\ii a\wh{k}}\;,
\label{eq:displace}
\end{equation}
and, as a consequence, the eigenvalues are not modified, while the
new eigenfunctions are easily computed as
\begin{eqnarray}
|t_{\pm}(k),D'\rangle & = & \e^{-\ii a\wh{k}}|t_{\pm}(k),D\rangle\nonumber \\
 & = & \frac{1}{\sqrt{2}}\left[\e^{-\ii ak}|k\rangle\pm \e^{\ii k(a+L)}|-k\rangle\right]\;.
\end{eqnarray}
It is straightforward to compute directly the action of $\widehat{T}_{D}$
in the wavenumber representation, 
\begin{equation}
\langle k|\widehat{T}_{D}|\psi\rangle=\frac{mL}{|k|\hbar}\left[\widetilde{\psi}(k)+\e^{-\ii kL}\frac{1}{kL}\sin\left({kL}\right)\widetilde{\psi}(-k)\right]\;,
\end{equation}
where $\widetilde{\psi}(k)=\la k|\psi\ra$. 
This allows us to study the functional aspects of the operator. In
particular, it is easy to check, using the requirement of normalizability,
that the domain of the operator is given by functions such that $\widetilde{\psi}(k)/k\rightarrow0$
as $k\rightarrow0$. The requirement of symmetry does not add further
limitations to the domain. As to the deficiency indices, they are
computed to be $(0,0)$. These computations are carried out for the
interval $D=[0,L]$, but, given the unitary equivalence of other intervals
of the same length, see Eq. (\ref{eq:displace}), the same results
carry over for regions composed of an arbitrary number of closed intervals.

It should come as no surprise that functions in the domain of the
(free particle) dwell-time operator must vanish at $p=0$ fast enough,
since the characteristic evolution of a generic wavefunction is dictated
by the free propagator, with its $t^{-1/2}$ temporal behavior. This
entails the divergence of the dwell time unless the state has no $p=0$ component
and the amplitude decay is faster (for
a simple analysis, see \cite{DEM02}). The divergence of the 
average dwell time for states with nonvanishing $p=0$ components also occurs for 
ensembles of classical particles.


To calculate the distribution $\Pi_\psi(t)$, see Eqs. (\ref{fomega}) and (\ref{eq:distfree}), we have 
in this case the characteristic function
\beqa
f_\psi(\omega)
& = & \int_{-\infty}^{\infty}\d k\, \e^{\ii\omega mL/(|k|\hbar)}\cos\left[\frac{\omega m}{k^2\hbar}\sin(kL)\right]\,|\widetilde{\psi}(k)|^{2}+
\nonumber\\
 &  & \, \ii\,\int_{-\infty}^{\infty}\d k\, \e^{\ii\omega mL/(|k|\hbar)}\sin\left[\frac{\omega m}{|k|k\hbar}\sin(kL)\right]
\quad \e^{-\ii kL}\widetilde{\psi}(k)\overline{\widetilde{\psi}(-k)}\;. 
\label{eq:transfdist}
\end{eqnarray}
The distribution $\Pi_{\psi}(t)$
has support only on the positive semi-axis, as can be seen from the
explicit expression for the eigenvalues. This, in turn, is a nontrivial check
of the correctness of the definition.

A further initially surprising result of this analysis is that for
highly monochromatic wavepackets (i.e., highly concentrated in one
point in the momentum representation), the probability density for
dwell times is generically bimodal because of the
two eigenvalues $t_{\pm}(k)$ expounded in Eq. (\ref{eq:freeeigenvalue}).
An experimental verification of the eigenvalues, and  of the quantum nature of the dwell time,  could be realized
with the aid of a  
matter-wave mirror located at a point $X>L$, reflecting an incident 
plane wave $|k\ra$.  
The resulting standing wave would take the form, up to a global phase factor,
\beq
|\psi_X\ra=|k\ra-\e^{-2\ii kX}|-k\ra\;.
\eeq
We may now compute the average dwell time between $0$ and $L$,
\beqa
\la \psi_X|\mathsf{T}|\psi_X\ra&=&\mathsf{T}_{kk}+\mathsf{T}_{-k-k}
-2\Re(\e^{-2\ii kX}T_{k-k})
\nonumber
\\
&=&2\frac{Lm}{|k|\hbar}
\left\{1-\cos[k(L+2X)]\frac{\sin(kL)}{kL}\right\}\;,
\eeqa
which oscillates between 
the maximum and minimum values $2t_\pm$, see (\ref{eq:freeeigenvalue});
they  occur at  
specific locations of the mirror, namely, 
\beqa
X_-&=&-\frac{L}{2}+\frac{\pi n}{k}\;,
\\
X_+&=&-\frac{L}{2}+\frac{\pi n}{k}+\frac{\pi}{2k}\;,
\eeqa
($n$ integer such that $X>L$), 
for which the standing wave $|\psi_X\ra$ becomes  proportional to
$|t_{\pm}\ra$. The proportionality factor $2^{1/2}$ accounts for the 
fact that the extrema correspond to {\it twice} the eigenvalues,
which is easy to interpret physically with reference to 
a corresponding classical scenario: the classical particle under similar circumstances would traverse the region twice, first rightwards and then leftwards. 
For any $X$ between the priviledged values $X_\pm$ given  above, $|\psi_X\ra$ is a linear superposition 
of the dwell time eigenvectors and thus the resulting average dwell time 
lies in a continuum between two times the eigenvalues.   
In a proposed experiment a highly monochromatic continuous beam 
would be sent towards the mirror and the particle density could be measured   
between $0$ and $L$ by fluorescence or other means. 
The oscillations of the signal as a function of $X$ would be in sharp contrast to the classical case, for which the the beam density and dwell time would remain unaffected by a change of the mirror's position.      
\subsection{A comparison with classicality}
We have already noticed some of the peculiarities of the quantum dwell time 
compared to the 
classical dwell time. Here we elaborate this comparison further.  
Consider a classical ensemble of free particles, described
by the initial probability density on phase space, $F(x,p)$. The
dwell-time distribution for this case is given by 
\begin{eqnarray}
\Pi_{\mathrm{class}}(t) & = & \int dx\, dp\,\delta\left(t-\frac{mL}{|p|}\right)\, F(x,p)
\nonumber \\
&=& \frac{mL}{t^{2}}\int dx\,\left[F\left(x,\frac{mL}{t}\right)
+F\left(x,\frac{mL}{-t}\right)\right]\;,
\end{eqnarray}
that is to say, the marginal momentum distribution evaluated at $mL/t$
and $-mL/t$, and multiplied by the normalization factor $mL/t^{2}$.
On the other hand, the distribution (\ref{eq:distfree}) obtained
from Eq. (\ref{eq:transfdist}) includes effects of interferences
between positive and negative momentum components, in two different
ways. In the first place, there is the obvious interference of the
last line of (\ref{eq:transfdist}); but, additional to this, the
argument of the cosine also reveals these effects. In order to see
better this point, let us examine a different operator,
\begin{equation}
\widehat{t}_{D}:=\widehat{\lambda}_+\widehat{T}_{D}\widehat{\lambda}_+
+\wh{\lambda}_-\wh{T}_D\wh{\lambda}_-=mL/|\wh{p}|\;,
\label{ted}
\end{equation}
where $\widehat{\lambda}_\pm$ are  the projectors onto the positive/negative momentum
subspaces. Notice that $\wh{\lambda}_\pm$ do not commute with $\widehat{T}_{D}$.
The last form in Eq. (\ref{ted}), specific of free motion, is particularly transparent and reproduces the
one of the classical dwell time. The eigenfunctions are $\ket{\pm k}$, $k>0$, and the corresponding eigenvalues are twofold degenerate and equal to the classical
time, $mL/\hbar|k|$. The distribution of dwell times for this operator,
and for positive-momentum states, is given by
\beq\label{pitau}
\pi_\psi(\tau) = \frac{mL}{\hbar \tau^2}  \left| \widetilde{\psi}\left(\frac{mL}{\hbar\tau}\right) \right|^2\;, 
\eeq
%
%
%
%
%
which coincides with the classical distribution for initial
wave functions whose support is limited to positive momenta. In this
manner we see that the distribution (\ref{eq:distfree}) incorporates
interferences even if only positive momenta are present. To be even
more explicit, observe that, for a state $\psi$ with only positive
momenta components, the second moment of the dwell-time
operator\index{dwell-time operator, second moment} (\ref{eq:definition})
is given by
\beqa
\left\langle \widehat{T}_D^2\right\rangle
&=&
\int_0^\infty \d k (\mathsf{T}^{\,2})_{kk}|\widetilde{\psi}(k)|^2=\int_0^\infty \d k (|\mathsf{T}_{kk}|^2+|\mathsf{T}_{k-k}|^2)|\widetilde{\psi}(k)|^2
\nonumber\\ &=&\int_{0}^{\infty}\d k\,\frac{m^2L^2}{k^2\hbar^2}\,\left[1+\frac{1}{k^2L^2}
\sin^2\left({kL}\right)\right]\left|\widetilde{\psi}(k)\right|^2\;,
\label{tkk2}
\eeqa
which contrasts with
\[
\left\langle \wh{{t}}^{\,\,2}_D\right\rangle =\int_{0}^{\infty}\d k\,\frac{m^2L^2}{k^2\hbar^2}\,\left|\widetilde{\psi}(k)\right|^2\,,\]
thus indicating that indeed the second term is due to quantum interference. 

As indicated above, there is another rather striking way of seeing
this point, by considering a highly monochromatic wavepacket. The classical
distribution would have a very sharply defined peak around $mL/p_{0}$,
where $p_{0}$ is the central momentum of the wavefunction, $p_0=k_0\hbar$; on the
other hand, the quantum distribution would have two sharp peaks,
centered on $t_{+}(k_{0})$ and $t_{-}(k_{0})$ respectively. 
The distance between peaks goes to zero as $p_{0}$
increases, as was to be expected since that is the classical limit.

The on-the-energy-shell version of $\widehat{t}_D$, $\mathsf{t}$, is also worth examining. By factoring out an energy delta function 
as in Eq. (\ref{dele}) we get for a plane wave $|k\ra$ the average 
$\mathsf{t}_{kk}=mL/(\hbar k)$, which is equal to $\mathsf{T}_{kk}$,  but the second moment differs, $(\mathsf{t}^2)_{kk}=(\mathsf{t}_{kk})^2=(\mathsf{T}_{kk})^2\le (\mathsf{T}^2)_{kk}$, see Fig. 1;
in other words, the
variance on the energy shell is zero since only one eigenvalue is possible for 
$\mathsf{t}$.
Contrast this with the extra term in Eq. (\ref{tkk2}), which again emphasizes the   
non-classicality of the dwell-time operator
$\widehat{T}_D$ and its quantum fluctuation.    
%
%
%
%
%
%
\section{The scattering case}
Let us now study the dwell time \index{dwell time in a potential barrier} in a potential barrier
or well without bound states, that is, the dwell time in an interval
which coincides with the finite support of the potential $V(x)$,
which presents no bound states. For simplicity, we shall assume as before 
that this interval starts at point $x=0$ and has length $L$. We shall
use the complete basis of incoming scattering stationary states, $\left|k^{+}\right\rangle $,
with the following position representation, for $k>0$, 
\begin{equation}
\left\langle x\big|k^{+}\right\rangle =\frac{1}{\sqrt{2\pi}}\left\{ \begin{array}{ll}
\e^{\ii kx}+R^{l}(k)\e^{-\ii kx} & \textrm{for }x<0\\
T^{l}(k)\e^{\ii kx} & \textrm{for }x>L\end{array}\right.\;,
\end{equation}
whereas,
for $k<0$, we have
\begin{equation}
\left\langle x\big|k^{+}\right\rangle =\frac{1}{\sqrt{2\pi}}
\left\{ \begin{array}{ll}
T^{r}(-k)\e^{\ii kx} & \textrm{for }x<0\\
\e^{\ii kx}+R^{r}(-k)\e^{-\ii kx} & \textrm{for }x>L\end{array}\right.
\end{equation}
(the superscripts $l,r$ in $T^{l}$, $R^{l}$, and $T^{r}$, $R^{r}$ stand for left
and right incidence).  
We have omitted any explicit expression in
the interval $[0,L]$ because of the potential dependence. The eigenvalues and eigenstates of the dwell-time operator can be formally computed using $\wh{H}\left|k^{+}\right\rangle =
(k^{2}\hbar^2/2m)\left|k^{+}\right\rangle$. 
To that end 
define
\begin{equation}
\xi(k)=\frac{\left\langle k^{+}\left|\chi_{D}(\wh{x})\right|k^{+}\right\rangle }{2\sigma(k)}\;,
\end{equation}
where
\begin{equation}
\sigma(k)=\left|\left\langle -k^{+}\left|\chi_{D}(\wh{x})\right|k^{+}\right\rangle \right|\;,
\end{equation}
and
\begin{equation}
\e^{\ii\varphi(k)}=\frac{\left\langle -k^{+}\left|\chi_{D}(\wh{x})\right|k^{+}\right\rangle }{\sigma(k)}\;.
\end{equation}
Additionally, for the sake of compactness in later formulae, let
\begin{equation}
\mu(k)=\frac{1}{2}\left[\xi(-k)-\xi(k)\right]\,.
\end{equation}
We then have that the eigenvalues of the dwell-time operator are
given by
\begin{equation}
t_{\pm}(k)=\frac{2\pi m\sigma(k)}{|k|\hbar}\,\left[\frac{\xi(k)+\xi(-k)}{2}\pm\sqrt{1+\mu^{2}(k)}\right]\;,
\end{equation}
while the eigenfunctions are 
\begin{equation}
|t_{\pm}(k)\rangle=N_{\pm}\left\{ \left|k^{+}\right\rangle +\e^{\ii\varphi(k)}\left[\mu(k)\pm\sqrt{1+\mu^{2}(k)}\right]\left|-k^{+}\right\rangle \right\}\;,
\end{equation}
and the normalization is given by
\begin{equation}
N_{\pm}=\frac{1}{\sqrt{2}}\left[1+\mu^2\pm\mu\sqrt{1+\mu^{2}(k)}\right]^{-1/2}\;.
\end{equation}
In fact, we can relate the quantities $\xi(k)$, $\sigma(k)$ and
$\varphi(k)$ to scattering data whenever the support of the scattering
potential is completely included in the region $D$. Let $\bar{\chi}_{D}(x)$
be the complementary function to $\chi_{D}(x)$. We can compute $\left\langle k'^{\,+}\right|\bar{\chi}(\wh{x})\left|k^{+}\right\rangle $
explicitly, and, using unitarity and the conditions this imposes on
the scattering amplitudes, we are led to the following expressions
(where we omit the arguments of the scattering amplitudes, since all
are evaluated at $k$, and denote derivative with respect to $k$
as $\partial_{k}$):
\begin{eqnarray*}
\left\langle k^{+}\left|\chi_{D}(\wh{x})\right|k^{+}\right\rangle  & = & \frac{L}{2\pi}\left|T^{l}\right|^{2}
+\frac{\ii}{2\pi}\left[R^{l}\partial_{k}\bar{R}^{l}+T^{l}\partial_{k}\bar{T}^{l}\right]
+\frac{\ii}{4\pi k}\left[\bar{R}^{l}-R^{l}\right]\,,
\\
\left\langle -k^{+}\left|\chi_{D}(\wh{x})\right|-k^{+}\right\rangle  & = & \frac{L}{2\pi}\left[1+\left|R^{r}\right|^{2}\right]
 +\frac{\ii}{2\pi}\left[R^{r}\partial_{k}\bar{R}^{r}+T^{r}\partial_{k}\bar{T}^{r}\right]
\nonumber\\
&+&\frac{\ii}{4\pi k}\left[\e^{-2\ii kL}\bar{R}^{r}-\e^{2\ii kL}R^{r}\right]\,,
\\
\left\langle -k^{+}\left|\chi_{D}(\wh{x})\right|k^{+}\right\rangle  & = & \frac{L}{2\pi}T^{l}\bar{R}^{r}
+\frac{\ii}{2\pi}\left[R^{l}\partial_{k}\bar{T}^{r}+T^{l}\partial_{k}\bar{R}^{r}\right]
+\frac{\ii}{4\pi k}\left[\bar{T}^{r}-\e^{2\ii kL}T^{l}\right]\,.
\end{eqnarray*}
Notice that the second term of each of the previous expressions can be identified
with the corresponding on-shell matrix elements of $-(\ii m/k\hbar)S^{\dagger}\partial_{k}S$.
That is, \index{Smith's delay time} with Smith's delay time!
\subsection{The Square Barrier}
\begin{figure}
\includegraphics[scale=.3]{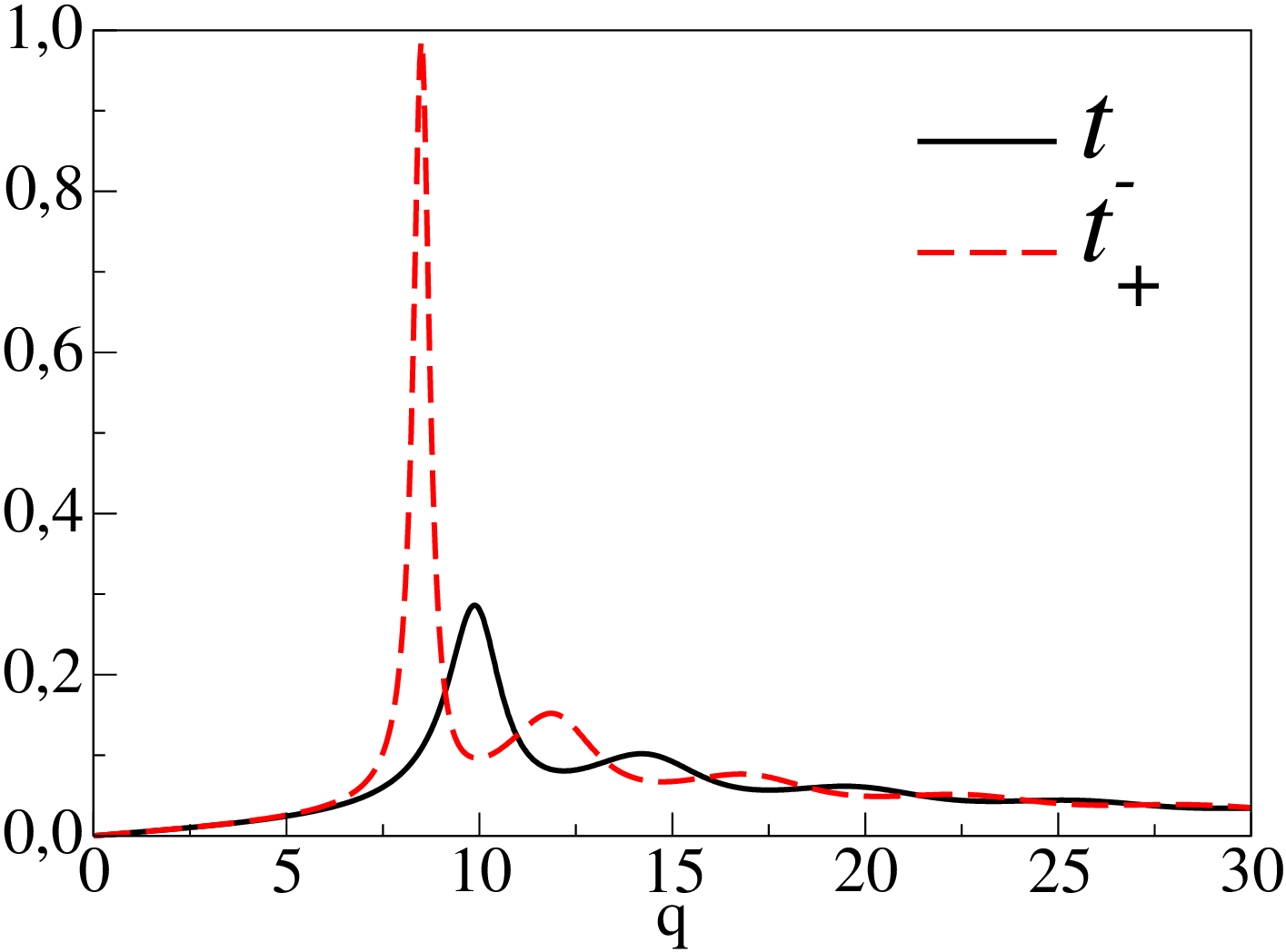}
\caption[]{Dwell time eigenvalues for the square barrier, in units of $mL^2/\hbar$, for $Q=8$} 
\label{figurebar}
\end{figure}

For the specific case of a square barrier, in which $V(x)=V_{0}[\theta(x-L)-\theta(x)]$,
with $\theta(x)$ being Heaviside's unit step function, we can put
to good use the symmetries of the Hamiltonian, namely $x\to L-x$
and $k\to-k$. These are realized on the scattering
states as
\[
\langle L-x|k^{+}\rangle=\e^{\ii kL}\langle x\left|-k^{+}\right\rangle \;,
\]
whence $\xi(k)=\xi(-k)$, and $\mu(k)=0$. Furthermore,
\begin{eqnarray*}
\left\langle k^{+}\left|\chi_{D}(\wh{x})\right|k^{+}\right\rangle   =  \frac{L\left|T(k)\right|^{2}}{2\pi \kappa^{2}}
\left\{ k^{2}-\frac{mV_{0}}{\hbar^2}\left[1+\frac{1}{2\kappa L}\sin\left({2\kappa L}\right)\right]\right\} \;,
\end{eqnarray*}
where $\kappa=\sqrt{k^{2}-2mV_{0}/\hbar^2},$ real and positive for positive above-the-barrier
momenta, and with positive imaginary part for tunneling momenta. After
some cumbersome algebra, one readily obtains
\begin{eqnarray*}
t_{\pm}(k)  =  \frac{\left|T(k)\right|^{2}mL}{2\hbar\left|k\right|\kappa^{2}}\left[k^{2}+\kappa^{2}\pm\left(\kappa^{2}-k^{2}\right)\cos{\kappa L}\right]
\left[1\pm\frac{1}{kL}\sin{kL}\right]\;,
\end{eqnarray*}
which is to be compared with (\ref{eq:freeeigenvalue}); namely,
the free particle result gets modulated by the transmission amplitude
and a potential dependent oscillatory factor. For high energies the
modulating term tends to one, and we recover the free particle case,
as one should. For small momenta, both eigenvalues tend linearly
to zero with the momentum, unlike the free case where one of them diverges.
It is also relevant that the dwell-time eigenvalues are
bounded above, and therefore the average value is also bounded! In
order to have a better grasp of this result, it is convenient to express
the eigenvalues in dimensionless terms
($k=q/L$, $V_{0}=\hbar^{2}Q^{2}/(2mL^{2})$
):
\[
t_{\pm}(k)=\frac{2mL^{2}}{\hbar}\frac{q\pm\frac{q}{\sqrt{{q^{2}-Q^{2}}}}\sin\sqrt{{q^{2}-Q^{2}}}}{2q^{2}-Q^{2}\pm Q^{2}\cos\sqrt{{q^{2}-Q^{2}}}}\;, 
\]
see Fig. \ref{figurebar}.
\subsection{Average dwell time from fluorescence measurements\label{sec6}}
In this subsection we shall model the measurement of the average 
dwell time $\tau_D=\la \wh{T}_D\ra$ of an ultracold atom at a square barrier created by a laser
shining perpendicularly to its motion with homogeneous intensity 
between $0$ and $L$. 

Consider a two level system coupled in a spatial region
to an off-resonance laser with large detuning,  
$\Delta\gg\gamma,\Omega$, where $\Delta$ is defined as the laser frequency minus 
the frequency of the atomic transition, $\gamma$ is the decay constant (inverse lifetime or Einstein's coefficient), and $\Omega$ is the Rabi frequency.

The amplitude for the atomic ground state up to the first 
photon detection is governed then  
by the following effective potential, see \index{complex potential} 
\cite{Heinzen,NEMH2003} or Chapter  ...,
\begin{equation}
V(x)=V_R- \ii V_I=\frac{\hbar \Omega^2}{4\Delta}-\ii\frac{\hbar\gamma
\Omega^2}{8\Delta^2}\;.
\label{pot}
\end{equation}
so that the average detection delay (life time of the ground state
if the atom at rest is put in the laser-illuminated region) is,
see e.g. \cite{ME08}, $4\Delta^2/\Omega^2\gamma$.
Whereas $\gamma$ is fixed for the atomic transition, 
$\Omega$ and $\Delta$ may  be controlled experimentally, and   
the ratio $\Omega^2/\Delta$ can always 
be chosen so that the real part of $V$ remains 
constant. This still leaves some freedom to fix their exact values 
which we may use to set the imaginary part.   
If we do so making sure that at most one fluorescence photon
is emitted per atom, i.e., 
$\tau_D\gg4\Delta^2/\Omega^2\gamma$, so that the fluorescence
signal produced by an atomic ensemble will be proportional to the 
absorption probability $A$, this signal provides, after calibration to take into 
the detector solid angle and efficiency, and approximation 
for the derivative (\ref{deriv}) and therefore to the average dwell time
for the potential (\ref{pot}) as 
\begin{equation}
\tau_D\approx\hbar A/(2V_{I})\;.
\label{approx}
\end{equation}
This follows by integrating 
$-dN/dt=(2 V_I/\hbar)\la\psi(t)|{\chi}_D(\wh{x})|\psi(t)\ra$
over time, $N$ being the surviving norm, and $A=1-N$ the absorption (fraction 
of atoms detected).
In the limit $V_I\to 0$ and for 
highly monochromatic incidence, the average (stationary) dwell time at the real potential is obtained,  
\begin{equation}
\mathsf{T}_{kk}=\lim_{V_I\to0}(\hbar/2)\partial_{V_I}A(k)\;,
\label{deriv}
\end{equation}
where $A(k)$ is the total absorption probability for incident wavenumber $k$.
The equivalence of this quantity with $(t_{+}(k)+t_{-}(k))/2$ can
readily be checked. 


\begin{figure}
\includegraphics[scale=.32]{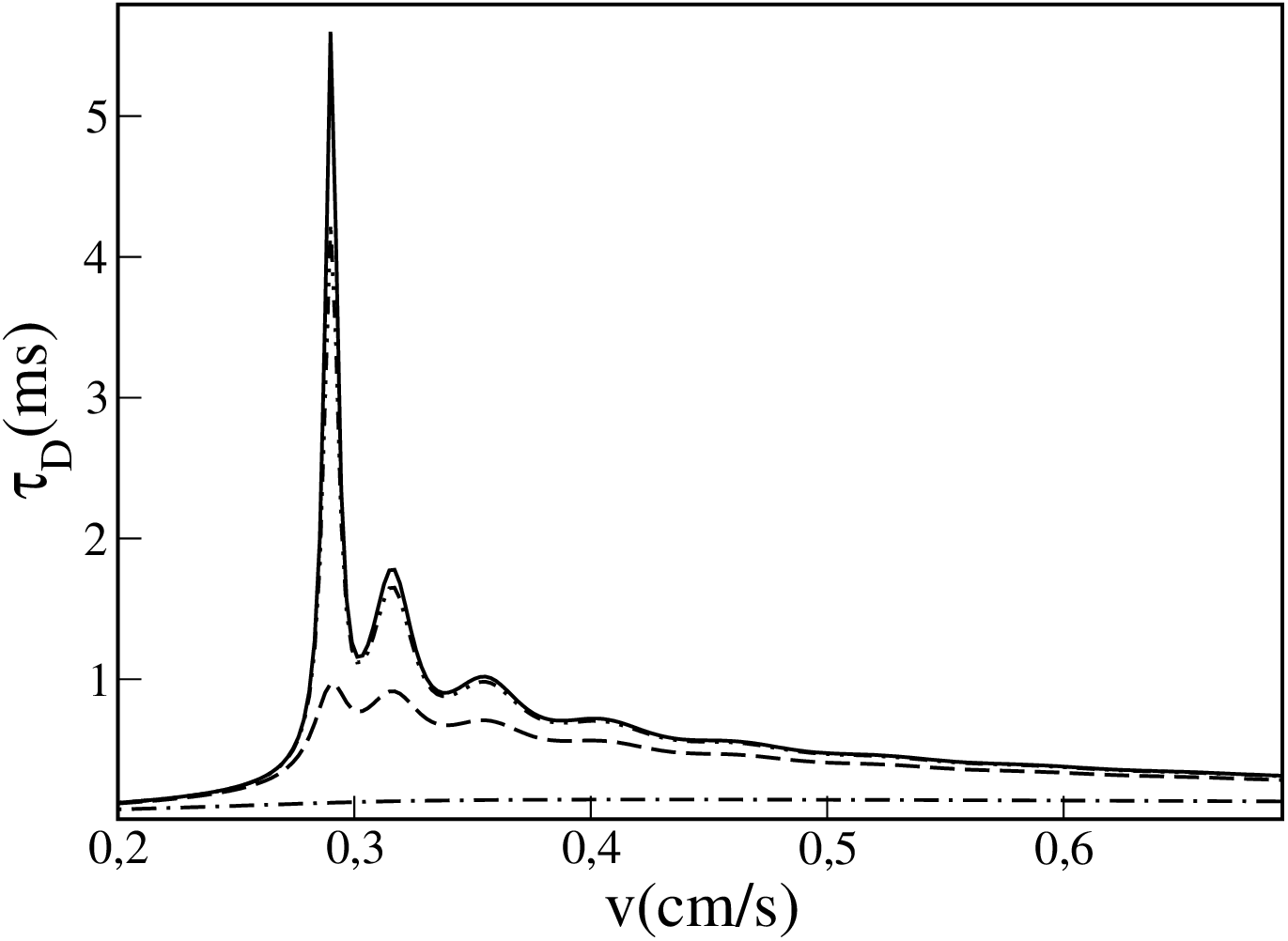}
\caption[]{Exact average dwell time (solid line) 
for Cs atoms crossing a square barrier of width 2$\mu$m
and height $8.2674\times 10^{3} s^{-1}\hbar$
(in velocity units, $0,28$ cm/s)
versus 
incident velocity. 
Approximations are calculated with Eq. (\ref{approx}) 
for $V_{I}=3.307 s^{-1}\hbar=V_1$
(indistinguishable from the exact result, $\Delta=2500\gamma$,
$\Omega=1.57\gamma$), 
10 $V_1$ (double dotted-dashed line, $\Delta=250\gamma$,
$\Omega=0.5\gamma$),
$10^2 V_1$ (dashed line, $\Delta=25\gamma$, $\Omega=.16\gamma$),
and $10^3 V_1$ (dotted dashed line, $\Delta=2.5\gamma$, $\Omega=0.05\gamma$).   
The transition
is at 852 nm with $\gamma$=33.3$\times10^6$s$^{-1}$;   
$\Delta$ and $\Omega$ are obtained from Eq. (\ref{pot})  
}
\label{figure1}
\end{figure}

%
%
%
%
%

One may think of relaxing the one-photon condition 
to get a proxy for the dwell time of an individual atom of the ensemble 
from the photons detected in an idealized  one-atom-at-a-time experiment.
For $V_R$ negligible versus $E$, it could be expected
that for some regime this distribution of emitted photons would also be
bimodal. 
For the bimodality to be observed, the characteristic interval between
modes ($\hbar/E$, where $E$ is the particle's energy) should be
greater than the characteristic interval between successive emission
of fluorescence photons, but 
these conditions and $\Delta>\gamma$ are  not compatible, and similar 
difficulties are found for on-resonance excitation.   
A pending task is the application of (deconvolution or operator normalization) techniques which have been successfully
applied to the arrival time, at least in theory.         
\begin{figure}
\includegraphics[scale=.3]{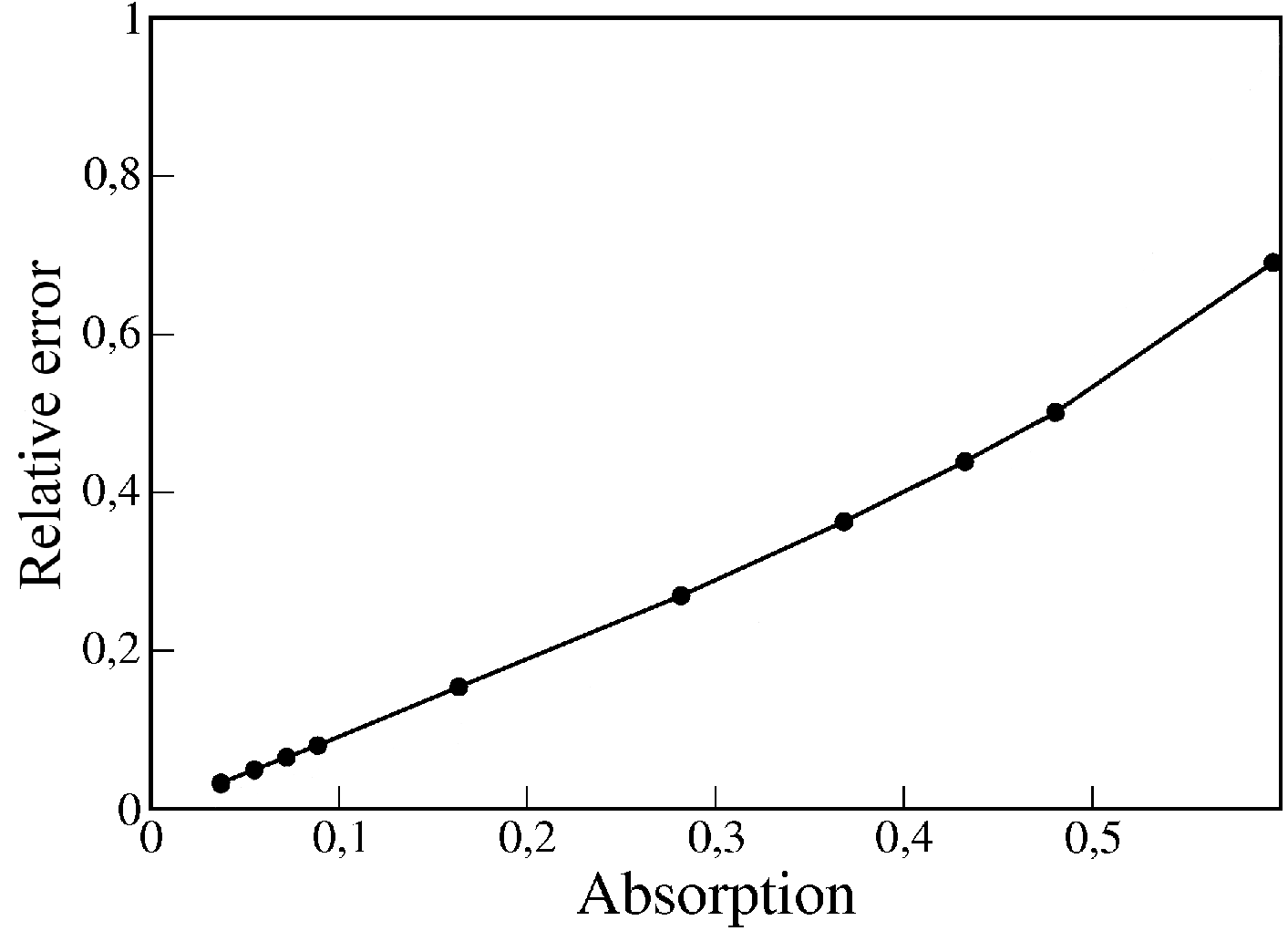}
\caption[]{Relative errors, calculated from the maxima 
of exact and approximate results, 
$|\tau_D(\rm{exact})-\tau_D(\rm{approx})|/\tau_D(\rm{exact})$,
versus 
the absorption (i.e., detection) probability 
used to calculate $\tau_D(approx)$ at the maximum. 
Same system as for the previous figure} 
\label{figure2}
\end{figure}

Figure \ref{figure1} shows the exact dwell time and approximations for 
several values of $V_{I}$ calculated for a transition of 
$Cs$ atoms (the details are in the figure caption).
A larger $V_{I}$ implies larger errors but also a stronger signal.
In practice the minimal signal requirements will determine the 
accuracy with which the dwell time can be measured. 
Figure \ref{figure2} shows the relative error of the dwell-time maxima 
versus the corresponding absorption probability.  
In this figures the beam is monochromatic.
We can check the reality of the quantum prediction
at hand, differing from the classical one; namely, that for all ingoing
waves the quantum mechanical dwell time is bounded, unlike the classical
one.  
    

%
%
%
\section{Some extensions\label{sec5}}
In this section we present miscellaneous extensions of  
the previous formalism and results:  
for bound states, for dwelling into 
states rather than in a spatial region, and for multiparticle systems.  
\subsection{The Harmonic oscillator and systems with bound states\label{sec:bounded}}
If we were to try the direct application of definition (\ref{eq:definition})
to the case of the harmonic oscillator we would soon run into trouble
due to divergent integrals. A natural option is to restrict the total 
time to one
period of the system, as follows: 
\begin{equation}
\widehat{T}_{D}=\int_{-T/2}^{T/2}\d \tau\, \e^{\ii\wh{H}\tau/\hbar}\chi_{D}(\wh{x})\, \e^{-\ii\wh{H}\tau/\hbar}\;,
\end{equation}
where the Hamiltonian is $\wh{p}^{\,2}/2m+m\omega^2\wh{x}^{\,2}/2$,
and $T=2\pi/\omega$.
We can thus write the dwell-time operator, restricted to one period,
as
\begin{equation}
\widehat{T}_{D}=T\sum_{n=0}^{\infty}\left\langle n\left|\chi_{D}(\wh{x})\right|n\right\rangle \left|n\right\rangle \left\langle n\right|\;.
\end{equation}
The eigenvalues, $T\left\langle n\left|\chi_{D}(\wh{x})\right|n\right\rangle $,
can be understood as the period times the proportion of that period
spent by the particle in the stationary state in the region $D$.
This suggests an alternative useful quantity for systems whose Hamiltonian
has a purely point spectrum. Define
\begin{equation}
\wh{\tau}_{D}=\lim_{T\rightarrow\infty}\frac{1}{T}\int_{-T/2}^{T/2}\d \tau\, \e^{\ii\wh{H}\tau/\hbar}\chi_{D}(\wh{x})\, \e^{-\ii\wh{H}\tau/\hbar}\;.
\end{equation}
Let the Hamiltonian be given by
\begin{equation}
\wh{H}=\sum_{n,\alpha}E_{n}\left|E_{n},\alpha\right\rangle \left\langle E_{n},\alpha\right|\;,
\end{equation}
with $\alpha$ a degeneracy index. Then it is easy to compute
\begin{equation}
\wh{\tau}_{D}=\sum_{n,\alpha,\beta}\left|E_{n},\alpha\right\rangle \left\langle E_{n},\alpha\right|\chi_{D}(\wh{x})\left|E_{n},\beta\right\rangle \left\langle E_{n},\beta\right|\;,
\end{equation}
that is to say, the operator that gives us the fraction of time spent
in region $D$ is diagonal in the energy basis, up to degeneracy in
energy.
%
%
%
%
%
\subsection{Times of Residence\label{tor}\index{time of residence}}
The construction above suggests an extension to the time
of residence, which is an analogue of the dwell time valid for systems in
which the concept of a region in space is not applicable. The question
we now purport to answer is the following one: how much time has a
system spent in a state $\left|\psi\right\rangle $? Or, alternatively,
from time 0 to $T$, what is the proportion of time the system has
spent in that state? Let then the projector associated with the state
$\left|\psi\right\rangle $ be denoted by $P$, and the unitary evolution
operator by $U(t)$. We define the time-of-residence operator \index{time of residence} as
\[
\wh{\tau}_{\psi}(0;T)=\int_{0}^{T}\d t\, U^{\dag}(t)PU(t)\;.
\]
For example \cite{S04}, consider a two-level system with
Hamiltonian 
\[
\wh{H}=\frac{\hbar}{2}\left(\begin{array}{cc}
0 & \Omega\\
\Omega & 0\end{array}\right)\;,
\]
and the state $\left|\psi\right\rangle =\left(\begin{array}{c}
1\\
0\end{array}\right)$. The time of residence gets written as\[
\wh{\tau}_{\psi}(0;T)=\left(\begin{array}{cc}
\frac{T}{2}+\frac{1}{2\Omega}\sin\left(\Omega T\right) & \frac{-\ii}{2\Omega}\left(1-\cos\left(\Omega T\right)\right)\\
\frac{\ii}{2\Omega}\left(1-\cos\left(\Omega T\right)\right) & \frac{T}{2}-\frac{1}{2\Omega}\sin\left(\Omega T\right)\end{array}\right)\;.
\]
This entails that the measurement of this quantity would inevitably
lead to one of the following two (eigen)values\index{time of residence, eigenvalues}:
\[
\tau_{\pm}=\frac{T}{2}\pm\frac{\sin\left(\Omega T/2\right)}{\Omega}\;.
\]
The fact that only two values can be obtained in each realization
of the experiment, and not a continuous distribution of time of presence
in the state $\left|\psi\right\rangle =\left(\begin{array}{c}
1\\
0\end{array}\right)$,
has been attributed to a predictive character of the measurement
involved, as opposed to an observation that extends over time through
the coupling with a weakly interacting clocking system \cite{S04}.

As a matter of fact, the average time spent by a particle prepared
at instant 0 in a generic state ranges from $\tau_{-}$ to $\tau_{+}$
for the eigenstate of $P$. Represent a generic pure or mixed state
$\rho$ by a vector in or on the Bloch sphere, $\rho=({1}+\vec{r}\cdot\vec{\sigma})/2$;
then the average time of residence in state $\left(\begin{array}{c}
1\\
0\end{array}\right)$ over the time interval $[0,T]$ is $T/2+\left[(1-\cos\left(\Omega T\right))r_{y}+\sin\left(\Omega T\right)r_{x}\right]/2\Omega$,
which, as stated, ranges over $[\tau_{-},\tau_{+}]$.

It should be noticed that the operator of time of residence from instant
0 to instant $T$ is generically not stationary; for the example at
hand we have
\[
\left[\wh{H},\wh{\tau}_{\psi}(0;T)\right]=\frac{\ii\hbar}{2}\left(\begin{array}{cc}
1-\cos\left(\Omega T\right) & i\sin\left(\Omega T\right)\\
-i\sin\left(\Omega T\right) & -1+\cos\left(\Omega T\right)\end{array}\right)\;.
\]
Notice that whenever $T=2n\pi/\Omega$, with $n$ a natural number,
the corresponding time of residence does indeed commute with the Hamiltonian;
this is in keeping with the expressions of Section \ref{sec:bounded}.


\subsection{Multi-particle systems}
The definition (\ref{eq:definition}) admits a straightforward extension
to systems of many particles, using the formalism of second quantization,
in which we perform the substitution
\begin{equation}
\label{nD}
\chi_{D}(\wh{x})\rightarrow \wh{n}_D=\int_{D}\d x\,\wh{a}_{x}^{\dagger}\wh{a}_{x}\,,
\end{equation}
where $\wh{a}_{x}$ is the annihilation operator at point $x$.
For free particles, that is to say, for a Hamiltonian of the form
\begin{equation}
\wh{H}=\int_{-\infty}^{\infty}\d k\,\frac{\hbar^2k^2}{2m}\wh{n}_{k}
=\int_{-\infty}^{\infty}\d k\,\frac{\hbar^2k^2}{2m}\wh{a}_{k}^{\dagger}\wh{a}_{k}\;,
\end{equation}
with $\wh{a}_{k}$ the operator that annihilates a particle with
wavenumber $k$, we can compute the dwell-time operator as
\begin{equation}
\widehat{T}_{D}=\int_{-\infty}^{\infty}\d k\,\frac{ mL}{|k|\hbar}\left[\wh{a}_{k}^{\dagger}\wh{a}_{k}+\frac{1}{kL}\e^{\ii kL}\sin\left({kL}\right)\wh{a}_{-k}^{\dagger}\wh{a}_{k}\right]\;.
\end{equation}
%

It is also feasible to write a dwell-time density operator, which
for the free case reads
\begin{eqnarray*}
\wh{\Pi}_{\mathrm{a}}(t) & = & \int_{-\infty}^{\infty}\d k\,\left\{ \frac{1}{2}\left[\delta\left(t-t_{+}(k)\right)+\delta\left(t-t_{-}(k)\right)\right]a_{k}^{\dagger}a_{k}\right.
\\
& & +\left.\frac{\e^{-\ii kL}}{2}\left[\delta\left(t-t_{+}(k)\right)-\delta\left(t-t_{-}(k)\right)\right]a_{k}^{\dagger}a_{-k}\right\} \;.
\end{eqnarray*}
It is easy to see that $\langle\psi|\wh{\Pi}_{\mathrm{a}}(t)|\psi\rangle$
reduces to the dwell-time density for the one particle marginal wavefunction.
The real multiparticle aspect of this construction will only be accessible
through dwell time - dwell-time correlation functions, for which a
suitable operational interpretation needs to be built. 
%
%
%
%


In addition, we note that for multiparticle systems, the alternative question might be posed, as to what is the time during which a given number of particles $n$ 
can be found in the region $D$. 
This naturally leads to the introduction of the dwell-time operator for $n$-particles \index{dwell-time operator, for $n$-particles}
\beqa
\widehat{T}_D(n)&=&\int_{-\infty}^{\infty} \d t\widehat{U}^{\dagger}(t)\delta_{\widehat{n}_D,n}\widehat{U}(t)\nonumber\\
&=&\frac{1}{2\pi}\int_{-\infty}^{\infty} \d t\int_{0}^{2\pi}\d\theta \widehat{U}^{\dagger}(t)
\e^{\ii\theta(\widehat{n}_D-n)}\widehat{U}(t)\;, 
\eeqa
where $\widehat{n}_D$ is the density operator restricted to the region $D$, defined in Eq. (\ref{nD}). 
Letting $\widehat{\rho}$ be the density matrix describing the state of the system, 
it is convenient to introduce the characteristic function,
\beqa
F(\theta;t)=\tr[\widehat{\rho}\widehat{U}^{\dagger}(t)\e^{\ii\theta\hat{n}_D}\widehat{U}(t)]\;,
\eeqa
whose Fourier transform is the atom number distribution in the region $D$ \cite{KKS00},
\beqa
P_D(n,t)=\frac{1}{2\pi}\int_{0}^{2\pi}\e^{-\ii n\theta}F(\theta;t)\d\theta\;. 
\eeqa
The meaning of $P_D(n,t)$ is precisely the probability for $n$ particles to be found in the spatial domain $D$ at time $t$.

%
Knowledge of the atom-number distribution can be used to compute the average dwell time for different number of particles, namely,
\beqa
\la \widehat{T}_D(n)\ra=\int_{-\infty}^{\infty} \d t\tr[\widehat{\rho}(t)\delta_{\widehat{n}_D,n}]
=\int_{-\infty}^{\infty} \d t P_D(n,t)\;.
\eeqa
%

\section{Relation to flux-flux correlation functions}

This section follows \cite{Munoz}, and examines   
a link between the dwell time and flux-flux correlation functions (ffcf) \index{flux-flux correlation functions, ffcc}    
that have been considered mostly in chemical physics to define reaction rates for microcanonical or canonical ensembles \cite{MST83}.
The motivation for this exercise is to relate the dwell-time distribution, and not just the average value, to other observables.  
%
%
%
%
%
%
%
%
%
%
\subsection{Stationary flux-flux correlation function}
%
%
Pollak and Miller \cite{PoMi-PRL-1984} have shown a connection between the average stationary dwell time and the first moment of  a ffcf. They define a quantum microcanonical ffcf $C_{PM}(\tau,k) = \tr\{\Re\, \wh{C}_{PM}(\tau,k) \}$ by means of the operator 
\beqa \label{eq:ffcf_PM}
\wh{C}_{PM}(\tau,k) &=& 2\pi\hbar [\wh{J}(x_2,\tau) \wh{J}(x_1,0) + \wh{J}(x_1,\tau) \wh{J}(x_2,0)
\nonumber\\
&-&\wh{J}(x_1,\tau) \wh{J}(x_1,0) - \wh{J}(x_2,\tau) \wh{J}(x_2,0)] \delta(E - \wh{H})\;, 
\eeqa
where $\wh{J}(x,t) = \e^{\ii \wh{H} t/\hbar} \frac{1}{2m} [\wh{p} \delta(\wh{x} - x) + \delta(\wh{x} -x) \wh{p}] \e^{-\ii \wh{H} t/\hbar}$ is the quantum mechanical flux operator in the Heisenberg picture, and $\wh{p}$ and $\wh{x}$  are the momentum and position operators.  

This definition is motivated from classical mechanics: Eq.~(\ref{eq:ffcf_PM}) counts flux correlations of particles entering $D$ through $x_1$ ($x_2$) and leaving it through $x_2$ ($x_1$) a time $\tau$ later. Moreover, particles may be reflected and may leave the region $D$ through its entrance point. This is described by the last two terms, where the minus sign compensates for the change of sign of a back-moving flux. Note that these negative terms lead to a self-correlation contribution that diverges for $\tau\to 0$.

We shall first derive the average correlation time and show its equivalence with the average dwell time. We shall only consider positive incident momenta, and define  $\wh{C}^+_{PM}$ by substituting $\delta(E-\wh{H})$ by  $\delta^+(E-\wh{H}):=\delta(E-\wh{H})\Lambda_+$, 
where $\Lambda_+$ is the projector onto the subspace of eigenstates of $H$ with positive momentum incidence. 
By means of 
the continuity equation,
\be \label{eq:cont}
-\frac{\d}{\d x} \wh{J}(x,t) = \frac{\d}{\d t} \wh{\rho}(x,t)\;,
\ee
where $\wh{\rho}(x,t) = \e^{\ii \wh{H} t/\hbar} \delta(\wh{x} - x) \e^{-\ii \wh{H} t/\hbar}$ is the (Heisenberg) density operator, 
$\wh{C}^+_{PM}(\tau,k)$ can be written as 
\beqa\label{eq:C_and}
\wh{C}^+_{PM}(\tau,k)
=- 2\pi\hbar \left( \frac{\d}{\d \tau} \wh{\chi}_D(\tau) \right)\left( \frac{\d}{\d t} \wh{\chi}_D(t) \right)_{t=0} \delta^+(E - \wh{H})\;.
\eeqa
By a partial integration and using the Heisenberg equation of motion the first moment of the Pollak-Miller correlation function is given by
\begin{equation}
\tr \left\{ \int_0^\infty \d \tau\, \tau \wh{C}^+_{PM}(\tau,k) \right\}
= \tr \left\{ 2\pi\hbar \int_0^\infty \d \tau \wh{\chi}_D(\tau) \frac{1}{i\hbar} [\wh{\chi}_D(0), \wh{H}] \delta^+(E - \wh{H})\right\}\;.
\end{equation}
Boundary terms of the form $\lim_{\tau\to\infty} \tau^{\gamma} \wh{\chi}_D(\tau), \gamma = 0,1,2$, are omitted here and in the following. 
The contribution of these terms should vanish when an integration over stationary wavefunctions is performed to account for the wavepacket dynamics, as it is done 
explicitly in the next section. 
For potential scattering the probability density decays generically as $\tau^{-3}$, which assures a finite 
dwell-time average, but for free motion it decays as $\tau^{-1}$ \cite{MDS95},
making 
$\tau_D$ infinite, unless the momentum wave function vanishes at $k=0$
sufficiently fast as $k$ tends to zero \cite{DEMN04}, as we have discussed
before.  
  
Writing the commutator explicitly and using the cyclic property of the trace gives
\beqa
&&\tr \left\{ \int_0^\infty \d \tau\, \tau \wh{C}^+_{PM}(\tau,k) \right\}
\nonumber\\
&=& \tr \left\{ 2\pi\hbar \int_0^\infty \d \tau\left( -\frac{\d}{\d \tau}\wh{\chi}_D(\tau) \right) \wh{\chi}_D(0) \delta^+(E - \wh{H})\right\}\;,
\eeqa
and integration over $\tau$ yields the final result,  
\beqa
\tr \left\{ \int_0^\infty \d \tau\, \tau \wh{C}^+_{PM}(\tau,k) \right\}
=2\pi\hbar \tr \left\{ \wh{\chi}_D(0) \delta^+(E - \wh{H}) \right\}
=\mathsf{T}_{kk}\;.
\eeqa
Expressing the trace in the basis $\ket{\phi_k}$ gives back the stationary dwell time of Eq.~(\ref{eq:dwell_stat}), i.e.\ the diagonal element of the on-the-energy-shell  dwell-time operator, $\mathsf{T}_{kk}$.

The
calculation of the average in \cite{PoMi-PRL-1984} is different in some respects: 
The coordinates $x_1$ and $x_2$ 
are taken to minus and plus infinity, but it can be carried out for finite 
values modifying Eq. (8) of \cite{PoMi-PRL-1984} accordingly; Formally there are no 
explicit boundary terms at infinity but  a regularization is required in Eq. 
(16) of \cite{PoMi-PRL-1984}, which is justified for wave packets; (c) $\delta(E-\wh{H})$
is used instead of $\delta^+(E-\wh{H})$. That simply provides an additional  contribution for negative momenta parallel to the one obtained here for positive momenta; 
(d) In our derivation the average correlation time is found to be real directly, even though $\wh{C}^+_{PM}(\tau,k)$ is not selfadjoint, whereas in 
\cite{PoMi-PRL-1984} the real part is taken. (The discussion of the imaginary time average in \cite{PoMi-PRL-1984} is based on a modified version of Eq.~(\ref{eq:ffcf_PM}).)

Next, we will show that the second moment of the Pollak-Miller ffcf equals the second moment of $\mathsf{T}$. This was not observed in Ref.~\cite{PoMi-PRL-1984}. Proceeding in a similar way as above, we start with   
\beq
{\cal I}=\tr\left\{\int_0^\infty \d \tau \tau^2 \widehat{C}^+_{PM}(\tau,k)\right\}\;.
\eeq
Integrating by parts twice, 
neglecting the term at infinity, using Heisenberg's equation of motion,
and the fact that $\phi_k$ is an eigenstate of $\wh{H}$, the real part is   
%
%
%
%
%
%
%
\beqa
\frac{{\cal I}+{\cal I}^*}{2}\!=\!\frac{2\pi m}{\hbar k}\!
\int_0^\infty\!\!\! \d \tau \la\phi_k|[\widehat{\chi}_D(\tau)\widehat{\chi}_D(0)
+\widehat{\chi}_D(0)
\widehat{\chi}_D(\tau)]|\phi_k\ra\;.
\eeqa
Introducing resolutions of the identity,   
\beqa
\Re {\cal I}&=&\!\Bigg\{\!\!\frac{2\pi m}{\hbar k}
\!\!\int_0^\infty\!\!\!\! \d \tau\!\!\int_{-\infty}^{\infty}\!\!\! \d k'
\!\!\int_{x_1}^{x_2}\!\!\! \d x\!\!\int_{x_1}^{x_2}\!\!\! \d x'
\e^{\ii (E-E')\tau/\hbar}
\phi_{k}^*(x)\phi_{k'}(x)\phi_{k'}^*(x')\phi_k(x')\!\Bigg\}
\nonumber\\
&+&c.c.\;,
\eeqa
where $c.c$ means complex conjugate. 
Making the changes $\tau\to-\tau$ and $x,x'\to x',x$ in the $c.c$-term, 
it takes the same form as the first one, but with the time integral from 
$-\infty$ to $0$. Adding the two terms, the $\tau$-integral provides an energy delta function that can be separated into two deltas which select $k'=\pm k$ to arrive at   
%
%
%
\beqa
\tr \left\{ \Re \int_0^\infty \d \tau\, \tau^2 \wh{C}^+_{PM}(\tau,k) \right\}
&=&\frac{4\pi^2 m^2}{\hbar^2 k^2}
\Bigg[\Big(\int_{x_1}^{x_2} \d x\, |\phi_k(x)|^2 \Big)^2
\nonumber\\
&+&\Big| \int_{x_1}^{x_2}\! \d x\, \phi_k^*(x) \phi_{-k}(x) \Big|^2 \Bigg]
=(\mathsf{T}^2)_{kk}\;.
\label{16}
\eeqa
%
In other words, the relation between dwell times and flux-flux correlation functions goes beyond average values and $C^+_{PM}(\tau,k)$ includes quantum features of 
the dwell time: note that the first summand in Eq. (\ref{16}) is nothing but $(\mathsf{T}_{kk})^2$, whereas the second summand is positive, which allows for a non-zero on-the-energy shell dwell-time variance
$(\mathsf{T}^2)_{kk}-(\mathsf{T}_{kk})^2$.  We insist that the stationary state considered has positive momentum, $\phi_k(x)$, $k>0$, but this second term  implies the degenerate partner $\phi_{-k}(x)$ as well, and is generically non-zero.

We shall see in the next section with a more general approach,
that these connections do not hold for higher moments.
%
%
%
%
\subsection{Time-dependent flux-flux correlation function}
%
%
%
%
%
%
%
A time-dependent version of the above flux-flux correlation function 
can be defined in terms of the operator
\beqa
\label{eq:C_op} 
\widehat{C}(\tau) &=& \int_{-\infty}^\infty \d t\, \bigl[\widehat{J}(x_2,t+\tau)
\widehat{J}(x_1,t) + \widehat{J}(x_1,t+\tau)
\widehat{J}(x_2,t)
\nonumber\\
&-& \widehat{J}(x_1,t+\tau)
\widehat{J}(x_1,t) - \widehat{J}(x_2,t+\tau)
\widehat{J}(x_2,t) \bigr]\;,
\eeqa
which leads to the flux-flux correlation function
\be \label{eq:Ctau_time}
C(\tau) = \langle \Re\,\wh{C}(\tau) \rangle_\psi\;,
\ee
where the real part is taken to symmetrize the non-selfadjoint
operator $\wh{C}(\tau)$ as before.

As in the stationary case, Eq.~(\ref{eq:Ctau_time}) counts flux correlations of particles entering $D$ through $x_1$ or $x_2$ at a time $t$ and leaving it either through $x_1$ or $x_2$ a time $\tau$ later. Moreover we integrate over the entrance time $t$. It can be shown that the first moment of the classical version of Eq.~(\ref{eq:C_op}), where $\widehat{J}$ is replaced by the classical dynamical variable of the flux, gives the average of the classical dwell time.

As in Eq.~(\ref{eq:C_and}), we may rewrite $\wh{C}(\tau)$ as
\be
\wh{C}(\tau) = -\int_{-\infty}^\infty \d t \frac{\d}{\d \tau} \wh{\chi}_D(\wh{x},t+\tau) \frac{\d}{\d t} \wh{\chi}_D(\wh{x},t)\;.
\ee
From here we note that the ffcf $C(\tau)$ is not normalized,  
\be
\int_0^\infty \d \tau\,\wh{C}(\tau) = \int_{-\infty}^\infty \d t\, \wh{\chi}_D(t) \frac{\d}{\d t} \wh{\chi}_D(t) = 0\;,
\ee
%
as a result of the self-correlation. 

Next we derive the average of the time-dependent correlation function. With a partial integration one finds
$$
\int_0^\infty \d \tau\,\tau \wh{C}(\tau) = \int_{-\infty}^\infty\! \d t   \int_0^\infty\! \d \tau\, \wh{\chi}_D(\wh{x},t+\tau)  \frac{\d}{\d t} \wh{\chi}_D(\wh{x},t)\;.
$$
%
A second partial integration with respect to $t$, replacing $\d/\d t$ by $\d/\d \tau$ and integrating over $\tau$ gives
\beqa
\label{eq:result_first}
\int_0^\infty \d \tau\,\tau \wh{C}(\tau) = \int_{-\infty}^\infty\! \d t\, \wh{\chi}_D^2(\wh{x},t) 
=\int_{-\infty}^\infty \d t\, \wh{\chi}_D(\wh{x},t) = 
\wh{T}_D\;,
\eeqa
where $\wh{\chi}_D^2=\wh{\chi}_D$ has been used. 
Eq.~(\ref{eq:result_first}) generalizes the result of Pollak and Miller to time-dependent dwell times.

A similar calculation can be performed for the second moment of $C(\tau)$. 
After three partial integrations with vanishing boundary contributions to get rid of the factor $\tau^2$ one obtains
$$
\int_0^\infty \d \tau\,\tau^2 \wh{C}(\tau) = 2 \int_{-\infty}^\infty\! \d t \int_0^\infty \d \tau\,\wh{\chi}_D(\wh{x},t+\tau)\wh{\chi}_D(\wh{x},t)\;.
$$
Making the substitutions $t+\tau \to t$ and $\tau \to -\tau$ in the complex conjugated term, we find
\begin{equation} \label{eq:result_second}
\mathrm{Re}\int_0^\infty \d \tau\,\tau^2 \wh{C}(\tau) = \wh{T}_D^2\;.
\end{equation}
%

\subsection{Example: free motion}
\begin{figure}
\includegraphics[scale=.35]{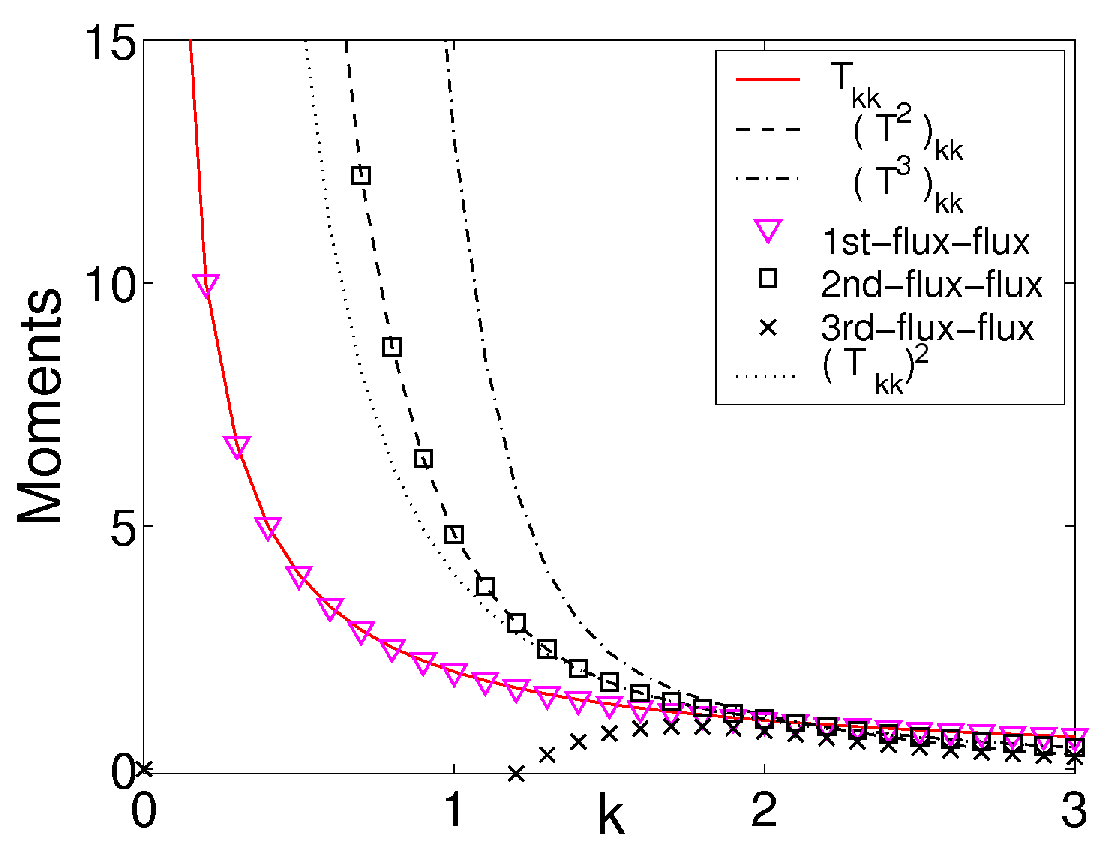}
\caption{Comparison of the first three moments: 
$\mathsf{T}_{kk}$, $(\mathsf{T}^2)_{kk}$ and $(\mathsf{T}^3)_{kk}$ (dotted-dashed line) with the corresponding moments
of the flux-flux correlation function, for a free-motion 
stationary state with fixed $k$. $(\mathsf{T}_{kk})^2$ is also shown (dotted line).  $\hbar = m = 1$ and $L = 2$}
\label{fig:moments}
\end{figure}

For a stationary flux of freely-moving particles with energy $E_k$, $k>0$, described by $\phi_k(x) = \braket{x}{k} = (2\pi)^{-1/2} \e^{\ii kx}$, the first three moments of the ideal dwell-time distribution on the energy shell 
are given by
\begin{eqnarray}
\mathsf{T}_{kk} &=& \frac{mL}{\hbar k}\;,\\
(\mathsf{T}^2)_{kk}&=& \frac{m^2L^2}{\hbar^2 k^2} \left(1 + \frac{\sin^2(kL)}{k^2 L^2} \right)\;,
\label{t2}\\
(\mathsf{T}^3)_{kk} &=& \frac{m^3L^3}{\hbar^3 k^3} \left(1 + 3 \frac{\sin^2(kL)}{k^2 L^2} \right)\;.
\end{eqnarray}
As proved above, the first two moments agree with the corresponding moments of the Pollak-Miller ffcf, but for the third moment we obtain instead
\beqa
\tr\left\{\Re\int_0^\infty \d \tau\, \tau^3 \wh{C}^+_{PM}(\tau,k)\right\}
= \frac{m^3L^3}{\hbar^3 k^3}\left[1 - \frac{3[1+ \cos^2(kL)]}{k^2 L^2} + \frac{3 \sin(2kL)}{L^3 k^3} \right]\;.
\label{28}
\eeqa
In Fig.~\ref{fig:moments} the first three moments are compared. The agreement between $(\mathsf{T}^3)_{kk}$ and Eq. (\ref{28}) is very
good for large values of $k$, but they clearly differ for small $k$. 
Nevertheless, the agreement of the first two moments suggests a similar behavior of $\Pi(\tau)$ and $C(\tau)$. 

To calculate $\Pi_\psi(\tau)$ for a wavefunction
$\widetilde{\psi}(k):=\la k|\psi\ra$
with only positive momentum components we use Eq. (\ref{eq:distfree}) 
and the explicit forms of the eigenvectors, Eq. (\ref{tvec})
and eigenvalues $t_\pm$, Eq. (\ref{eq:freeeigenvalue}),    
%
\be \label{eq:dwell_distri_free}
\Pi_\psi(\tau) = \frac{1}{2} \sum_j \sum_{\gamma = \pm} \frac{|\widetilde{\psi}(k_j^\gamma(\tau))|^2}{|F'_\gamma(k_j^\gamma(\tau))|}\;,
\ee
where the $j$-sum is over the solutions $k_j^\gamma(\tau)$ of the equation $F_\gamma(k) \equiv t_\gamma(k)-\tau = 0$
and the derivative is with respect to $k$.

\begin{figure}
\includegraphics[scale=.35]{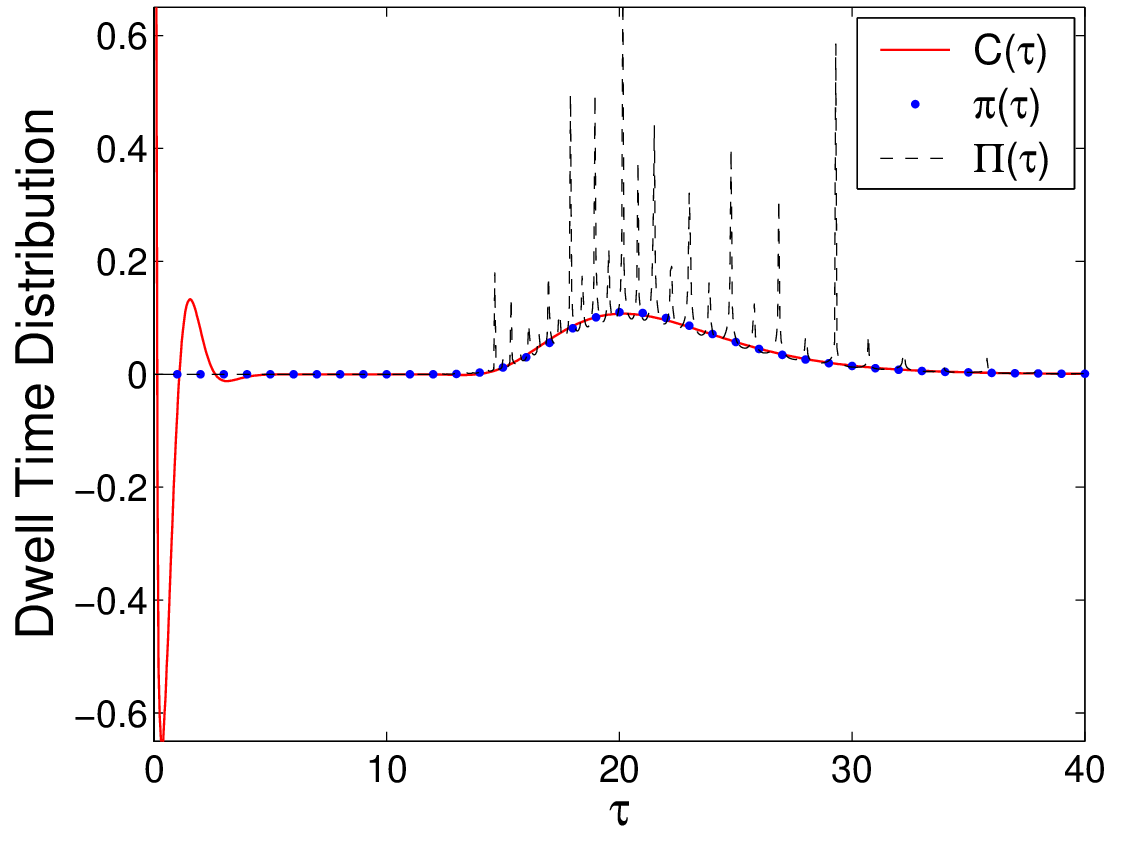}
\caption{Comparison of dwell-time distribution $\Pi(\tau)$ (dashed line) and flux-flux correlation function $C(\tau)$ (solid line) for the freely moving wave packet (\ref{eq:k_distri}). Furthermore, the alternative free-motion dwell-time distribution $\pi(\tau)$, Eq. (\ref{pitau}), 
is plotted (circles). 
We set $\hbar=m=1$
and $|x_0|$ large enough to avoid overlap of the initial state
with the space region $D=[0,45]$. $k_0=2$, $\Delta k =0.4$, and $\alpha = 0.5$}
\label{fig:Ctau1}
\end{figure}

We use the following wavefunction \cite{charac},
\be \label{eq:k_distri}
\widetilde{\psi}(k) = {\cal N}(1-\e^{-\alpha k^2})\,\e^{-(k-k_0)^2/[4(\Delta k)^2]} \e^{-\ii kx_0}\Theta(k)\;,
\ee
where ${\cal N}$ is the normalization constant and $\Theta(k)$ the step function.
For the free flux-flux correlation function we write
\be
C(\tau) = \Re \int_0^\infty \d k \int_0^\infty \d k'\, \widetilde{\psi}^*(k) \widetilde{\psi}(k') \bra{k} \wh{C}(\tau) \ket{k'}\;,
\ee
and $C_{kk'}(\tau) = \bra{k} \wh{C}(\tau) \ket{k'}$ in the free case 
\beqa
C_{kk'}(\tau) = \frac{m}{2\pi \hbar k} \delta(k-k') \frac{\d^2}{\d\tau^2} \Big[2g(\hbar k\tau/m) 
- g(\hbar k\tau/m-L)-g(\hbar k\tau/m+L)\Big]\;,
\eeqa
where 
\be
g(x) = -2\e^{\ii mx^2/(2\hbar\tau)} \left(\frac{\ii\pi\hbar\tau}{2m} \right)^{1/2} + \ii\pi x\, \mt{erfi} \left(\sqrt{\frac{\ii m}{2\hbar \tau}}\,x \right)\;.
\ee
The result is shown in Fig.~\ref{fig:Ctau1}. The ffcf shows a hump around the mean dwell time but it oscillates for small $\tau$ and diverges for $\tau\to 0$. As discussed above, this is due to the self-correlation contribution of wavepackets which are at $x_1$ or $x_2$ at the times $t$ and $t+\tau$ {\it without} changing the direction of motion in between. A similar feature has been observed in a traversal-time distribution derived by means of a path integral approach \cite{Fertig-PRL-1990}. 

In contrast, $\Pi_\psi(\tau)$ behaves regularly for $\tau\to 0$, but shows peaks in the region of the hump. This is because the denominator of Eq.~(\ref{eq:dwell_distri_free}) becomes zero if the slope of the eigenvalues $t_\pm(k)$ is zero, which occurs at every crossing
of $t_+(k)$ and $t_-(k)$.

The distribution $\pi_\psi(\tau)$, see Eq. (\ref{pitau}), is also computed:   
it 
agrees with $C(\tau)$ in the region near the average dwell time and it 
tends to zero for $\tau\to 0$. However, it does not show the resonance peaks of $\Pi_\psi(\tau)$.

In absence of a direct dwell-time measurement, the physical significance $\wh{T}_D$ and $\wh{t}_D$
depends on their relation to other observables. 
The present results indicate that the second moment of the 
flux-flux correlation function is related to the former and not to the later, 
providing indirect support for the physical relevance of the dwell-time 
resonance peaks, but other observables could behave differently.    
%
\subsection{Approximations}
\begin{figure}
\includegraphics[scale=.3]{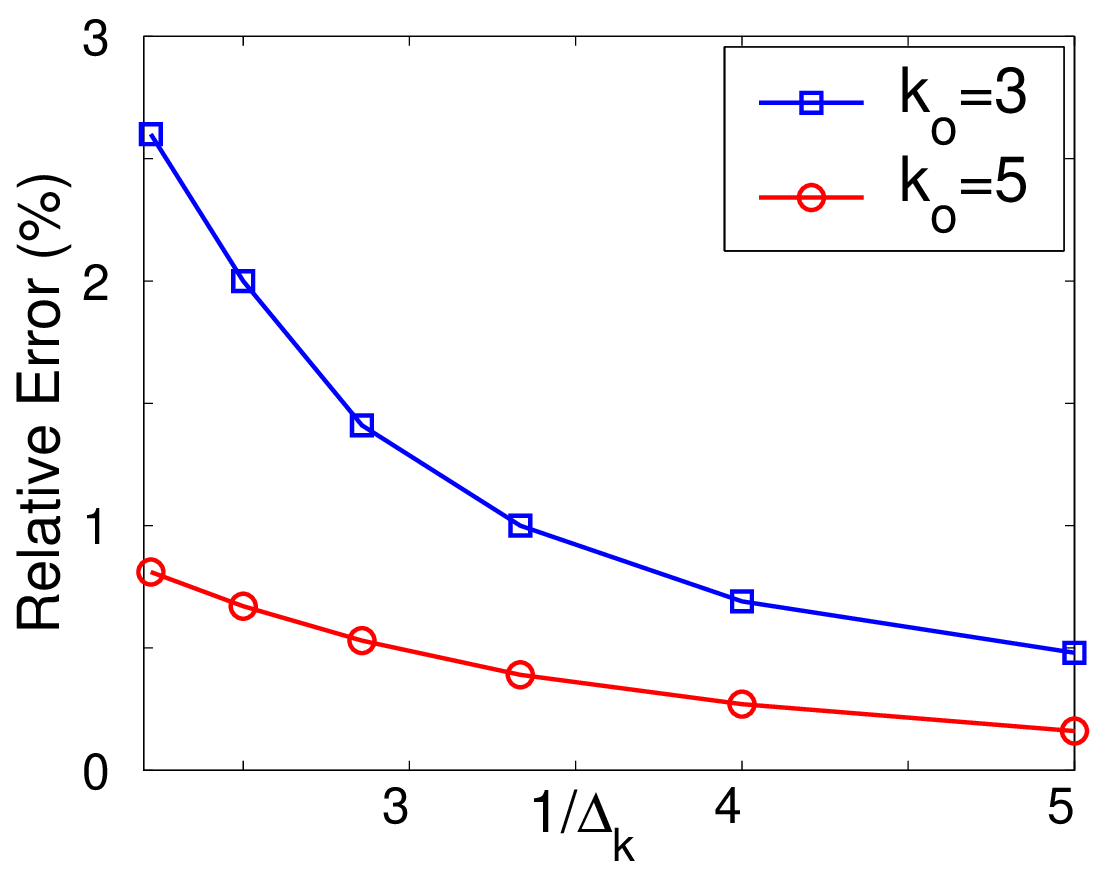}
\caption{Comparison of the relative error of $\langle \wh{T}_D^2\rangle$ using the approximation ${C_{0}} (\tau)$ instead of $C(\tau)$ for free motion.
$\alpha=0.5$, $\hbar = m = 1$ and $L = 100$}
\label{fig:delta}
\end{figure}

%
%
%
%
%
While the previous results bring dwell-time information closer to experimental realization, the difficulty is translated onto 
the measurement of the ffcf, not necessarily an easy task. 
A simple approximation is to substitute the expectation of the 
product of two flux operators by the product of their  
expectation values, which are the current densities. 
Using the wave packet of Eq. (\ref{eq:k_distri}), we have compared the times obtained 
with the full expression (\ref{eq:Ctau_time}) 
and with this approximation in Fig. \ref{fig:delta}. 
The two results approach each other as $\Delta_k \to 0$, also   
%
%
%
by increasing $L$
and/or $k_0$.

The exact result can be approached systematically, still making 
use of ordinary current densities, as follows:   
First decompose  
$\widehat{J}(x_i,t+\tau)\widehat{J}(x_j,t)$  
by means of 
\beqa \label{eq:unit}
\widehat{1}&=&\widehat{P}+\widehat{Q}\;,
\\
\widehat{P}&=&\ket{\psi}\bra{\psi}\;,
\eeqa
%
%
%
so that 
%
%
%
\beqa 
\label{eq:jj}
\widehat{J}(x_i,t+\tau)\widehat{J}(x_j,t)
=\widehat{J}(x_i,t+\tau)(\ket{\psi}\bra{\psi}+\widehat{Q})\widehat{J}(x_j,t)\;.
\eeqa
It is useful to decompose $\widehat{Q}$ further in terms of a basis of states 
orthogonal to $\ket{\psi}$ and to each other, $\{\ket{\psi_j^Q}\}$, 
%
%
%
\be
\label{eq:q_explic}
\widehat{Q}=\sum_j\ket{\psi_j^Q}\bra{\psi_j^Q}\;,
\ee
%
%
%
that could be generated by means of a Gram-Schmidt orthogonalization procedure.  
Now we can split Eq. ({\ref{eq:C_op}), 
\be\label{eq:C_op_1y2} 
\widehat{C}(\tau) = \widehat{C}_0(\tau)+\widehat{C}_1(\tau)\;, 
\ee
where $\widehat{C}_0(\tau)$ has the structure of $\widehat{C}$, but with 
$P$ inserted between the two flux operators in each of the four terms.
Similarly $\widehat{C}_1(\tau)$ has $Q$ inserted
and can be itself decomposed using Eq. (\ref{eq:q_explic}).    
%
%
%

We define 
$C(\tau)=C_0(\tau)+C_1(\tau)$ by taking the real 
real part of $\la \psi|\wh{C}_0(\tau)+\wh{C}_1(\tau)|\psi\ra$. 
$C_0$ is the zeroth order approximation discussed before and only involves 
ordinary, 
measurable current densities \cite{DEHM2002}.  
The non-diagonal terms from $C_1$,  
$\bra{\psi}\widehat{J}(x_i,t)\ket{\psi_j^Q}\bra{\psi_j^Q}\widehat{J}(x_j,t+\tau)\ket{\psi}$	can also be related to diagonal elements of $\widehat{J}$
by means of the  
auxiliary states
\beqa
\ket{\psi_1}&=&\ket{\psi}+\ket{\psi_j^Q}\;,
\nonumber
\\
\ket{\psi_2}&=&\ket{\psi}+\ii\ket{\psi_j^Q}\;,
\nonumber
\\
\ket{\psi_3}&=&\ket{\psi}-\ii\ket{\psi_j^Q}\;,  
\eeqa
since  
\beqa
\label{eq:pol_met}
\bra{\psi}\widehat{J}(x,t)\ket{\psi_j^Q}&=&\frac{1}{2}\bra{\psi_1}\widehat{J}(x,t)\ket{\psi_1}
-\frac{1}{4}\bra{\psi_2}\widehat{J}(x,t)\ket{\psi_2}
\nonumber\\
&-&\frac{1}{4}\bra{\psi_3}\widehat{J}(x,t)\ket{\psi_3}
+\frac{\ii}{4}\bra{\psi_3}\widehat{J}(x,t)\ket{\psi_3}
\nonumber\\
&-&\frac{\ii}{4}\bra{\psi_2}\widehat{J}(x,t)\ket{\psi_2}\;.
\eeqa
\section{Final comments}
The quantum 
dwell-time distribution of a particle in a spatial region, and its second moment present non-classical features even for 
the simplest case of a freely moving particle, such as bimodality (due to 
two different 
eigenvalues for the same energy) with the strongest deviations from classical 
behavior 
occurring for de Broglie wavelengths of the order or larger than the region
width. Progress in ultracold atom manipulation makes plausible the observation of these effects, and motivates further effort to achieve an elusive direct 
measurement of the dwell times, or to link the distribution and its moments to other   observables. We have in this regard pointed out that the flux-flux correlation 
function provides access to the second moment.  The potential impact in cold atom time-frequency metrology \cite{sei07,mou07}, 
and other fields in which the dwell time 
plays a prominent role (such as conductivity \cite{Bu}, or chaos) remains an open question.

\section*{Acknowledgements}
We acknowledge discussions  with M B\"uttiker, J. A. Damborenea and B. Navarro. 
This work has been supported by Ministerio de Educaci\'on y Ciencia (FIS2006-10268-C03-01) and the Basque Country University (UPV-EHU, GIU07/40). 
%
%

\end{document}